\documentclass[aps,a4paper,twocolumn,superscriptaddress,amsmath,amssymb,floatfix,10pt]{revtex4-2}
\usepackage[T1]{fontenc}
\usepackage{newtxtext,newtxmath}
\usepackage{mathrsfs}
\usepackage{graphicx,booktabs,microtype,siunitx,xcolor,array}
\usepackage{capt-of} 
\usepackage[colorlinks=true,linkcolor=blue!60!black,citecolor=blue!65!black,urlcolor=blue!60!black]{hyperref}
\usepackage{orcidlink}
\hypersetup{
  pdftitle={Parity-dependent wave absorption and optical characteristics of a cored magnetic black hole},
  pdfauthor={Mohsen Fathi and Faizuddin Ahmed},
  pdfsubject={EPJC presubmission manuscript; magnetic black holes, nonlinear electrodynamics, optical characteristics, coupled wave absorption}
}
\newcommand{\F}{\mathcal F}
\newcommand{\LL}{\mathcal L}

\newcommand{\ee}{\mathrm e}
\newcommand{\gmode}{\mathrm g}

\begin{document}
\raggedbottom

\title{Parity-dependent wave absorption and optical characteristics of a cored magnetic black hole}

\author{Mohsen Fathi\,\orcidlink{0000-0002-1602-0722}}
\email{mohsen.fathi@ucentral.cl}
\affiliation{Centro de Investigaci\'on en Ciencias del Espacio y F\'isica Te\'orica (CICEF), Universidad Central de Chile, La Serena 1710164, Chile}

\author{Faizuddin Ahmed\,\orcidlink{0000-0003-2196-9622}}
\email{faizuddinahmed15@gmail.com}
\affiliation{Department of Physics, The Assam Royal Global University, Guwahati-781035, Assam, India}

\date{\today}

\begin{abstract}
We study light propagation and coupled gravitational--electromagnetic waves around a black hole supported by a nonlinear magnetic field. The same field action fixes the geometry, the mass--charge relation, and the wave potentials. We calculate two critical impact parameters for light and solve the coupled scattering equations in both parity sectors. In eight solutions of one action, the two optical scales differ by $3.6$--$7.2\%$. The wave absorption also depends on parity. For $\ell=2$, the gap between the larger absorption eigenvalues reaches about $9.0$ percentage points. A wider frequency scan reveals a stronger effect in the second eigenchannel. The difference between the two absorption traces reaches $24.3$ percentage points. The trace equals twice the absorption of an equal incoherent mixture of incoming gravitational and electromagnetic fields. A controlled test of the wave operator shows that the nonlinear electromagnetic factor $\kappa$ plays an important role in the parity gap. We also examine four $\ell=3$ cases and use charge-matched Reissner--Nordstr\"om black holes as a separate control. The optical and wave calculations provide two complementary properties of the same magnetic source.
\end{abstract}
\maketitle

\begin{figure*}[tbp]
\centering
\includegraphics[width=.98\textwidth]{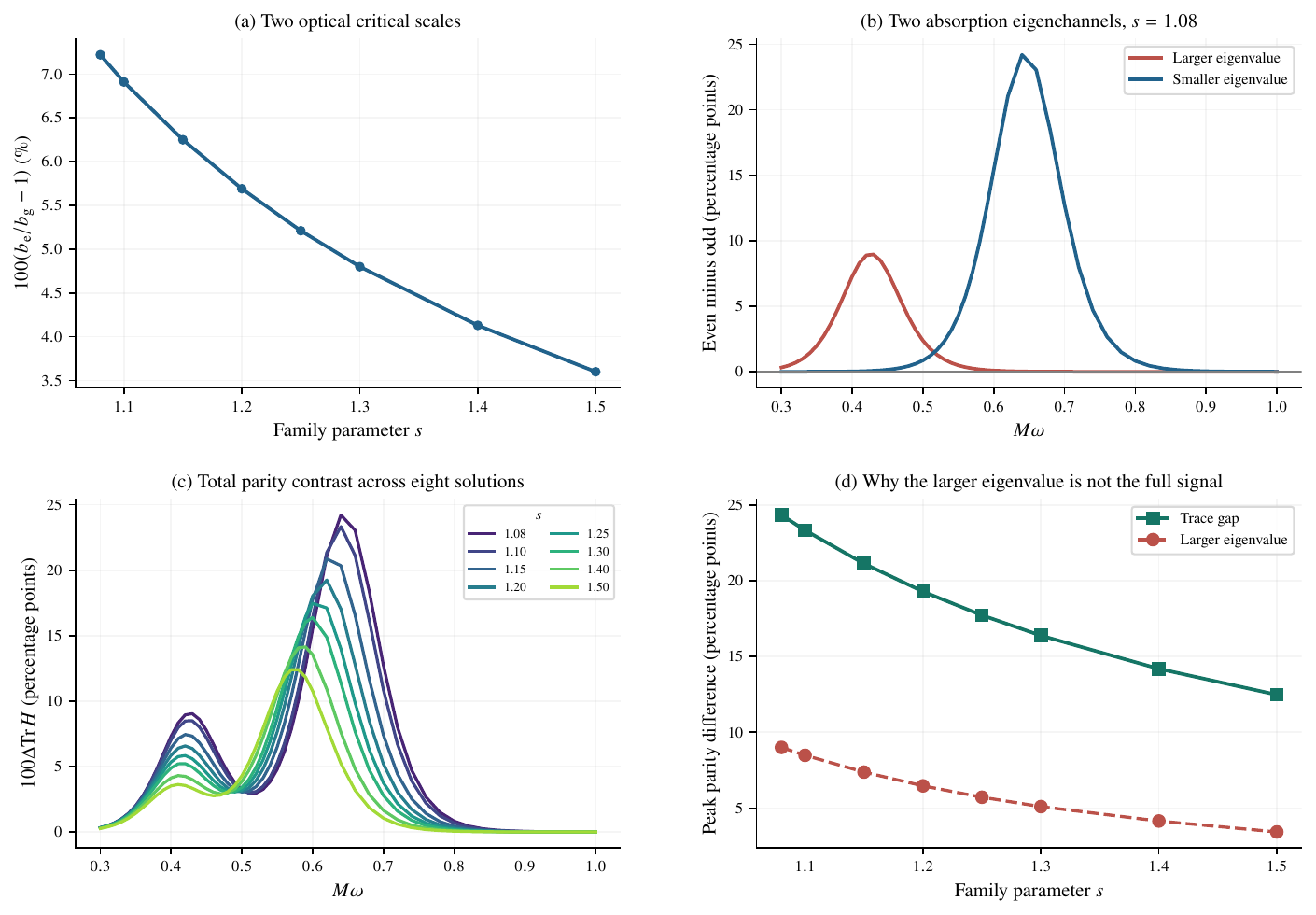}
\caption{A common magnetic source produces two optical scales and a parity-dependent wave response. (a) Separation of the geometric and extraordinary critical impact parameters. (b) Differences between the two absorption eigenvalues at $s=1.08$. The larger eigenvalue saturates first. (c) Differences of the absorption trace for all eight solutions. (d) Refined trace peaks compared with the peaks of the larger eigenvalue. The wave data are real-frequency solutions of the coupled equations. The absorption differences are in percentage points.}\label{fig:optical_wave_fingerprint}

\end{figure*}

\section{Introduction}
Black holes give us a way to test how matter changes strong gravity. A metric tells us how ordinary null rays move. If the source is a nonlinear electromagnetic field, the field itself can also change the propagation of electromagnetic disturbances \cite{Novello2000,Moreno2003,Walia2024}. We therefore ask a simple question: \emph{how do light propagation and coupled gravitational--electromagnetic wave absorption respond to the same magnetic source?}

Our model is a static magnetic black hole. Its density is smooth near the center and has a power-law tail at large radius. The field action reduces to Maxwell theory in the weak-field limit. We derive the relation between mass, charge and core scale from this action. We then use the same family of solutions to study photon trajectories and coupled wave absorption. The central profile is hidden behind the horizon; the optical and wave signals come from the exterior.

Our main result is a difference in absorption between the two metric-parity sectors of the same black hole. There are two absorption eigenvalues in each sector. The larger one gives a clear gap near $M\omega=0.43$, while the smaller one carries a stronger signal at higher frequencies. We compare both eigenvalues, their sum, and specified incoming gravitational and electromagnetic waves. Figure~\ref{fig:optical_wave_fingerprint} gives the main result. The comparison with Reissner--Nordstr\"om (RN) is a separate control.

The main ingredients have a long history. Regular magnetic solutions were studied in early Einstein--nonlinear-electrodynamics work \cite{AyonBeato1998,Bronnikov2001}. Coupled perturbations and parity splitting were developed in Refs.~\cite{ReggeWheeler1957,Zerilli1970,Moreno2003,Chaverra2016,Nomura2022}. Two optical characteristics and related light rings have also been studied \cite{Novello2000,DePaula2026}. Our contribution is a calculated response for this particular Maxwell-limit magnetic model, with one fixed action, an explicit optical comparison, and flux-normalized wave scattering across a controlled sequence of black holes.

There are useful nearby studies. The generalized Hayward magnetic model has been examined with coupled gravitational--electromagnetic perturbations \cite{Pradhan2026,Daghigh2022}. Electromagnetic quasinormal modes in a different asymptotically AdS nonlinear-electromagnetic background provide another useful contrast \cite{NLEDQNM2026}. Einasto-based geometric models and their perturbations have also been discussed \cite{Konoplya2026,Lutfuoglu2026,Bolokhov2026}; their matter prescriptions differ from ours. Rotating perfect-fluid dark matter backgrounds lead to a different class of photon-orbit and image calculations \cite{Nashed2026}. These works give context for our results but do not identify the same source or parameter family.

We also compare the absorption with charge-matched RN black holes. The mass and charge of each RN control are fixed by the asymptotic metric; no spectral parameter is fitted. The RN comparison is a separate check, while the direct parity gap remains our main wave result. The optical critical scales, wave absorption and astronomical emission-ring measurements are different quantities \cite{Gralla2019,Johnson2020,EHTpolar2021,Perlick2022}. We keep that distinction when discussing possible observations.

\section{Matter reconstruction and gravitational background}\label{sec:matter}
In units $G=c=M=1$, let
\begin{equation}\label{eq:metric}
\begin{aligned}ds^2&=-f(r)dt^2+\frac{dr^2}{f(r)}+r^2d\Omega^2,\\ f(r)&=1-\frac{2m(r)}{r}.\end{aligned}
\end{equation}
The Einstein equations for a static anisotropic source with $p_r=-\rho$ give
\begin{equation}\label{eq:stress}
 m'=4\pi r^2\rho,\quad p_r=-\rho,\quad p_t=-\rho-\frac r2\rho'.
\end{equation}
The exact Einasto profile is $\rho_{\rm E}=\rho_0\exp[-(r/a)^q]$, with $a,q>0$. For a magnetic monopole with $\F\equiv F_{\mu\nu}F^{\mu\nu}/4=Q^2/(2r^4)$, a single-invariant nonlinear-electrodynamic (NLED) background completion has $\LL(\F(r))=\rho(r)$. Its kinetic coefficient is
\begin{equation}
 \LL_{\F}=-\frac{r\rho'(r)}{4\F(r)}.
\end{equation}
For the exponential profile, $\LL_{\F}$ vanishes as $r\to\infty$, faster than any power. Its extraordinary characteristic coefficient
\begin{equation}\label{eq:kappa-def}
 \kappa=1+\frac{2\F\LL_{\F\F}}{\LL_{\F}}=-\frac12\left(3+\frac{r\rho''}{\rho'}\right)
\end{equation}
becomes $\kappa_{\rm E}=[q(r/a)^q-q-2]/2$ and diverges in the far field. Maxwell theory instead requires $\LL_{\F}\to1/(4\pi)$ and $\kappa\to1$. This is a no-go for this \emph{specified} magnetic reconstruction of an exact exponential density, not for all possible halo or field theories. Regular magnetic completions, reverse-engineered actions and their physical consistency conditions have an extensive literature \cite{Balart2014,Bokulic2024,Bokulic2022,Bokulic2026,Mazharimousavi2026}.

\subsection{From the Einstein equations to a single-invariant magnetic source}
We obtain the geometry and its matter source from the same field equations. We adopt standard Einstein gravity, spherical symmetry and a magnetic monopole. The magnetic invariant is positive and falls as $r^{-4}$. For the chosen Lagrangian convention, a stress tensor derived from the nonlinear electromagnetic action has $T^t{}_t=T^r{}_r=-\rho$. This equality fixes the radial equation of state but does not by itself fix the tangential pressure or the kinetic response of fluctuating electromagnetic fields. Those depend on the derivative of the action. The radial and angular Einstein equations give Eq.~(\ref{eq:stress}), and the conservation equation $\nabla_\mu T^{\mu r}=0$ supplies a useful consistency check: its single nontrivial component is $p_r'+2(p_r-p_t)/r=0$. Inserting $p_r=-\rho$ reproduces $p_t=-\rho-r\rho'/2$ exactly. The pressure relation is consequently a property of the sourced geometry, not an extra phenomenological ansatz.

For a magnetic configuration, $\mathcal F=Q^2/(2r^4)$ is monotone outside the center. A prescribed differentiable radial density may therefore be used to define an electromagnetic action \emph{along this solution} by eliminating $r$ between $\rho(r)$ and $\mathcal F(r)$. This is a convenient inverse-construction method, also explored systematically in the regular-black-hole literature \cite{Balart2014,Bokulic2024}. One must not confuse it with a proof that the resulting action has a well-posed Cauchy problem everywhere or that the same action supports an arbitrary two-parameter set of masses and magnetic charges. The reconstructed Lagrangian is a function of an invariant, so its derivatives determine how perturbations see the source. This point distinguishes an Einstein-consistent completion from a metric-only deformation used for test rays.

To expose the inverse map directly, differentiate $\mathcal F(r)$: $\mathcal F'=-4\mathcal F/r$. From $\rho'=\mathcal L_{\mathcal F}\mathcal F'$ we obtain
\begin{equation}\label{eq:inverse-LF}
 \mathcal L_{\mathcal F}=-\frac{r\rho'}{4\mathcal F}.
\end{equation}
This relation fixes the field response from the density profile. The Maxwell limit $\mathcal L_{\mathcal F}\to1/(4\pi)$ requires a density tail proportional to $r^{-4}$ with the appropriate charge normalization. In particular, it cannot be inferred from an asymptotically Schwarzschild metric alone. A mass profile may yield finite total mass and regular central curvature while its constitutive field theory fails the standard electromagnetic weak-field limit.

The exact Einasto exponential illustrates this distinction. Let $y=(r/a)^q$. Then $-r\rho_{\rm E}'/\rho_{\rm E}=q y$, and Eq.~(\ref{eq:inverse-LF}) gives $\mathcal L_{\mathcal F}\propto r^{q+4}e^{-y}$ at large $r$. This tends to zero, not to $1/(4\pi)$. The associated characteristic coefficient calculated below also grows without bound at large $r$. These are precise statements about this monopole reconstruction in a single-invariant NLED theory; they do not rule out a collisionless Einasto distribution coupled through a different stress tensor, a two-invariant electromagnetic action, or a multicomponent source. The model built here is chosen deliberately because the observable electromagnetic propagation is part of the question.

\subsection{A Maxwell-complete magnetic reconstruction and its exterior domain}
We use a profile that is exponential to leading order at small radius but has an electromagnetic tail:
\begin{equation}\label{eq:profile}
 \boxed{\rho(r)=\rho_0\left[1+\left(\frac rR\right)^q\right]^{-4/q}},\quad 1\le q\le2.
\end{equation}
With $a=R(q/4)^{1/q}$, its central expansion is
\begin{equation}\label{eq:inner}
 \log(\rho/\rho_0)=-(r/a)^q+\frac q8(r/a)^{2q}+O(r^{3q}).
\end{equation}
Equation~(\ref{eq:inner}) shows the exponential behavior near the center. The outer tail is different. In the wave sequence the horizon lies at about $4$--$7$ times $R$, while the optical and wave calculations use the region outside it. Thus the results describe the exterior of a cored magnetic black hole.

 Define $x=(r/R)^q$ and $B_q=B(3/q,1/q)$. Normalizing the total mass to unity gives
\begin{equation}\label{eq:mass}
 \rho_0=\frac{q}{4\pi R^3 B_q},\qquad m(r)=I_{x/(1+x)}\left(\frac3q,\frac1q\right),
\end{equation}
where $I_z$ is the regularized incomplete beta function. This yields $m\propto r^3$ near the center and finite curvature scalars. The density is finite at $r=0$; we do not assert that every derivative is analytic there for noninteger $q$.

At infinity, the mass function satisfies
\begin{equation}\label{eq:rn}
 f(r)=1-\frac2r+\frac{Q^2}{r^2}+O(r^{-2-q}),\quad Q^2=\frac{2qR}{B_q}.
\end{equation}
The field source is explicit. Put $\F_0=Q^2/(2R^4)$; since $x=(\F_0/\F)^{q/4}$,
\begin{equation}\label{eq:lagrangian}
 \LL(\F)=\rho_0\left[1+(\F_0/\F)^{q/4}\right]^{-4/q}.
\end{equation}
For small $\F$, the action has the expansion $\LL(\F)=\F/(4\pi)+O(\F^{1+q/4})$. Its leading term is Maxwell theory. The next term can have a nonanalytic second derivative at $\F=0$ for the stated $q$ range. We use the magnetic branch $\F>0$. The monopole field satisfies its NLED field equation, while Eq.~(\ref{eq:stress}) gives the Einstein-source relation. Each chosen $(q,R)$ specifies a \emph{different} action and its associated charge. Consequently our matched-metric sweep is a comparison \emph{across theories}, not a population of independently adjustable states of one fixed NLED theory.

\begin{figure}[tbp]
\centering\includegraphics[width=\columnwidth]{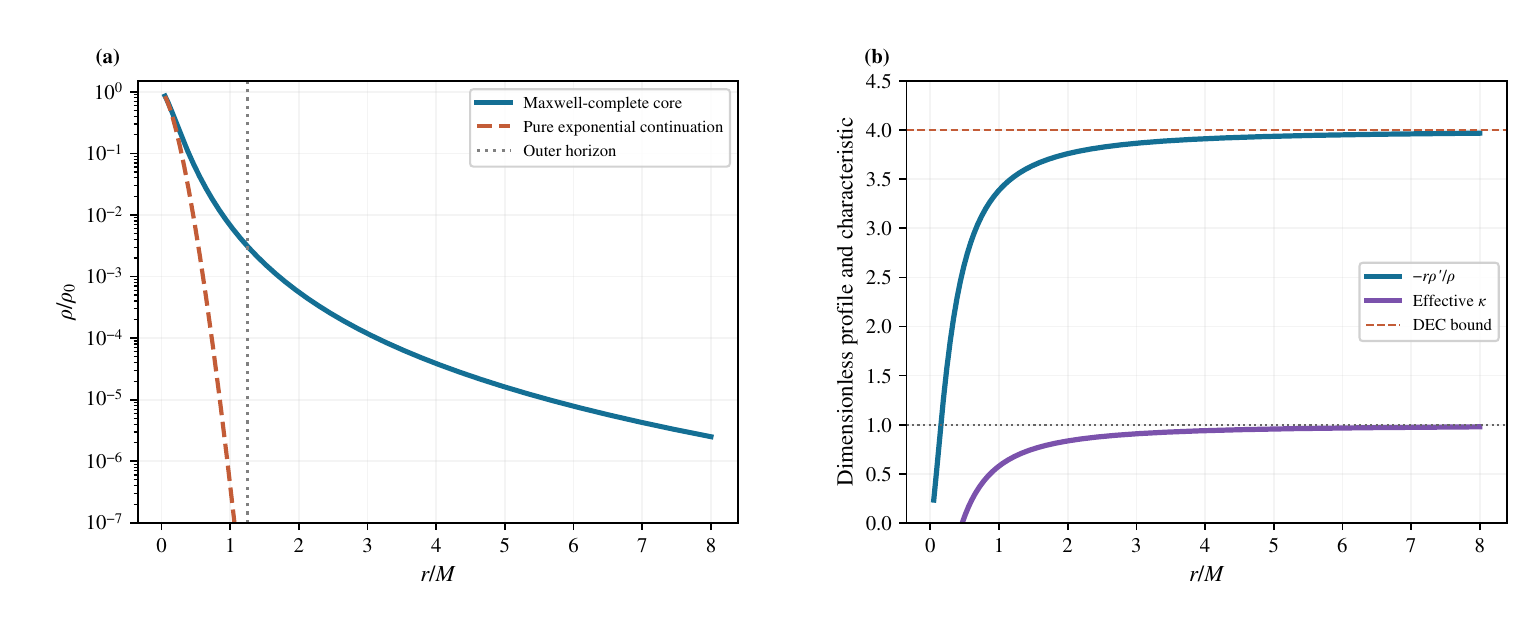}
\caption{(a) The Maxwell-complete profile and an exponential profile with the same central leading term, illustrated for $q=1.5$. Their outer tails differ. (b) The effective characteristic factor and the logarithmic density slope. The density slope remains below the dominant-energy-condition bound of four.}
\label{fig:completion}
\end{figure}

The energy conditions and characteristics are unusually transparent (Fig.~\ref{fig:completion}). Setting $S\equiv-r\rho'/\rho=4x/(1+x)$, one has $0\le S<4$. Hence $\rho\ge0$, $\rho+p_r=0$, $\rho+p_t=S\rho/2\ge0$, and $\rho-p_t=(4-S)\rho/2\ge0$. Both the weak and dominant energy conditions hold at every positive radius. Substitution into Eq.~(\ref{eq:kappa-def}) gives
\begin{equation}\label{eq:kappa}
 \boxed{\kappa(r)=1-\frac{q+4}{2(1+x)}}.
\end{equation}
Thus $\kappa\to1$ as $r\to\infty$, and the extraordinary cone is Lorentzian where $x>(q+2)/2$. We check that this inequality holds \emph{everywhere outside the outer horizon} for all black holes used below. At small radii the extraordinary sector can lose hyperbolicity, and we do not infer global well-posedness or stability of the coupled system.

\subsection{Mass normalization, core geometry and the RN limit}
The normalization used in Eq.~(\ref{eq:mass}) can be derived without fitting an integration constant to the horizon. Starting from the mass integral,
\begin{equation}
 M=4\pi\rho_0\int_0^\infty\!\frac{r^2\,dr}{[1+(r/R)^q]^{4/q}},
\end{equation}
we put $t=(r/R)^q$ and then $z=t/(1+t)$. The integral becomes $(4\pi\rho_0R^3/q)B(3/q,1/q)$, yielding the first equality of Eq.~(\ref{eq:mass}) when $M=1$. The incomplete version with an upper limit $r$ gives the regularized beta function in the second equality. This derivation supplies a practical computational cross-check: the mass obtained by direct quadrature of $4\pi r^2\rho$ must agree with the incomplete-beta expression at the same $(q,R,r)$, including near the outer horizon where small relative errors can move a photon-orbit root.

The central and asymptotic expansions give complementary physical information. Near the center $\rho=\rho_0[1-(4/q)(r/R)^q+O(r^{2q})]$. Integrating the density shows $m=(4\pi\rho_0/3)r^3+O(r^{3+q})$ and hence
\begin{equation}\label{eq:deSitter-core}
 f(r)=1-\frac{8\pi\rho_0}{3}r^2+O(r^{2+q}).
\end{equation}
This is a regular de Sitter-like core at the level of curvature invariants. It does not imply analyticity of all higher radial derivatives for fractional $q$, nor does it prove stability at the center. These caveats matter because a regular \emph{background} metric need not be a dynamically stable object in a nonlinear electromagnetic theory \cite{Bokulic2026,DeFelice2025}.

At $r\gg R$, expand the density rather than the metric directly:
\begin{equation}\label{eq:rho-asymptotic}
 \rho=\frac{\rho_0R^4}{r^4}\left[1-\frac4q\left(\frac Rr\right)^q+O(r^{-2q})\right].
\end{equation}
The mass deficit $M-m(r)=\int_r^\infty4\pi u^2\rho(u)du$ consequently equals $4\pi\rho_0R^4/r+O(r^{-1-q})$. The coefficient of $1/r^2$ in $f$ is therefore $8\pi\rho_0R^4=Q^2$, which agrees with Eq.~(\ref{eq:rn}). This relation ties the gravitational far field to the magnetic charge inferred from the electromagnetic action. It removes the freedom to choose an RN comparison charge independently when we compute wave absorption.

\subsection{Constitutive derivatives, energy inequalities and hyperbolicity}
Energy inequalities constrain the background stress tensor, while hyperbolicity constrains the principal propagation law of perturbations. For the profile $S\equiv-r\rho'/\rho=4x/(1+x)$, the tangential pressure is $p_t=\rho(S/2-1)$. It follows that
\begin{equation}
 \rho+p_t=\frac{2x}{1+x}\rho,\qquad
 \rho-p_t=\frac{2}{1+x}\rho.
\end{equation}
Both expressions are nonnegative for every $x\ge0$, while $\rho+p_r=0$. Therefore the weak and dominant energy conditions are satisfied for this source. Their satisfaction does not necessarily ensure causal electromagnetic characteristics; that is an independent property of derivatives of $\mathcal L$. Similar distinctions motivate causality analyses of charged NLED solutions \cite{Tomizawa2023,RussoTownsend2026}.

Writing the action in terms of $y=(\mathcal F_0/\mathcal F)^{q/4}=x$ gives its exact derivative
\begin{equation}\label{eq:LF-full}
 4\pi\mathcal L_{\mathcal F}=\left(\frac{x}{1+x}\right)^{1+4/q}.
\end{equation}
At small $\mathcal F$, $x\to\infty$ and $4\pi\mathcal L_{\mathcal F}\to1$. At fixed magnetic charge $x\to0$ towards the center, and the kinetic coefficient tends to zero. The extraordinary angular characteristic coefficient can be written in two equivalent ways, $\kappa=1+2\mathcal F\mathcal L_{\mathcal F\mathcal F}/\mathcal L_{\mathcal F}$ or $\kappa=1-(q+4)/[2(1+x)]$. Both can be evaluated independently from the source profile to check algebra and signs.

A simple analytical boundary for the extraordinary characteristic sector follows from $\kappa>0$:
\begin{equation}\label{eq:kappa-threshold}
 r>r_\kappa\equiv R\left(\frac{q+2}{2}\right)^{1/q}.
\end{equation}
We do not require $\kappa$ to stay positive all the way to $r=0$, since the physical scattering boundary problem is posed outside the event horizon. Instead, we test $r_h>r_\kappa$ for each sampled black-hole solution and monitor the constitutive coefficients throughout the exterior integration domain. This is a sharper statement than calling the entire spacetime ``causal'' on the basis of an exterior check. It also explains why the regular center and the exterior wave problem should be discussed separately, as in studies of kinetic and gradient instabilities \cite{NomuraYoshidaSoda2020,DeFelice2025}.

\section{Optical observables and strong-field ray propagation}\label{sec:optics}
For a magnetic $\LL(\F)$ model, geometric and extraordinary polarization rays follow different characteristic Hamiltonians \cite{Moreno2003,Tomizawa2023}:
\begin{equation}\label{eq:U}
 U_{\gmode}(r)=\frac{f(r)}{r^2},\qquad U_{\ee}(r)=\frac{f(r)\kappa(r)}{r^2}.
\end{equation}
A circular unstable ray lies at $U_j'(r_j)=0$ and has $b_j=U_j(r_j)^{-1/2}$, for $j\in\{\gmode,\ee\}$. The corresponding asymptotic critical impact parameters are meaningful here because $\kappa\to1$ and $f\to1$ at large radius.

We fit $R$ independently for 21 values $q=1,1.05,\ldots,2$ so that \emph{every geometric branch} obeys $b_{\gmode}=4M$. All sampled solutions possess an outer horizon, a unique exterior unstable maximum for each branch, $\kappa>0$ on the exterior and a decaying characteristic barrier. Table~\ref{tab:matched} contains representative results. The second critical impact parameter changes from $4.514M$ to $4.236M$, although the ordinary geometric critical impact remains fixed at $4M$. This translates to a polarization-dependent split of $12.86\%$--$5.91\%$. It is a background-field optical prediction of the stated action, not a mock observational constraint. The complete 21-model comparison is given in Table~\ref{tab:app-matched-21}.

\begin{table}[tbp]
\caption{Matched geometric $b_{\gmode}=4M$; all entries have $M=1$. $\bar a_j$ is the logarithmic strong-deflection coefficient in Eq.~(\ref{eq:strong}).}
\label{tab:matched}
\centering\small
\begin{tabular}{@{}rrrrrr@{}}\toprule
$q$ & $R/M$ & $r_h/M$ & $b_{\ee}/M$ & $\bar a_{\gmode}$ & $\bar a_{\ee}$ \\
\midrule
1.0 & 0.2005 & 1.3012 & 4.5142 & 1.2427 & 1.1327\\
1.5 & 0.3201 & 1.2484 & 4.3488 & 1.3021 & 1.2043\\
2.0 & 0.4031 & 1.2071 & 4.2363 & 1.3434 & 1.2495\\
\bottomrule\end{tabular}
\end{table}

\begin{figure}[tbp]
\centering\includegraphics[width=\columnwidth]{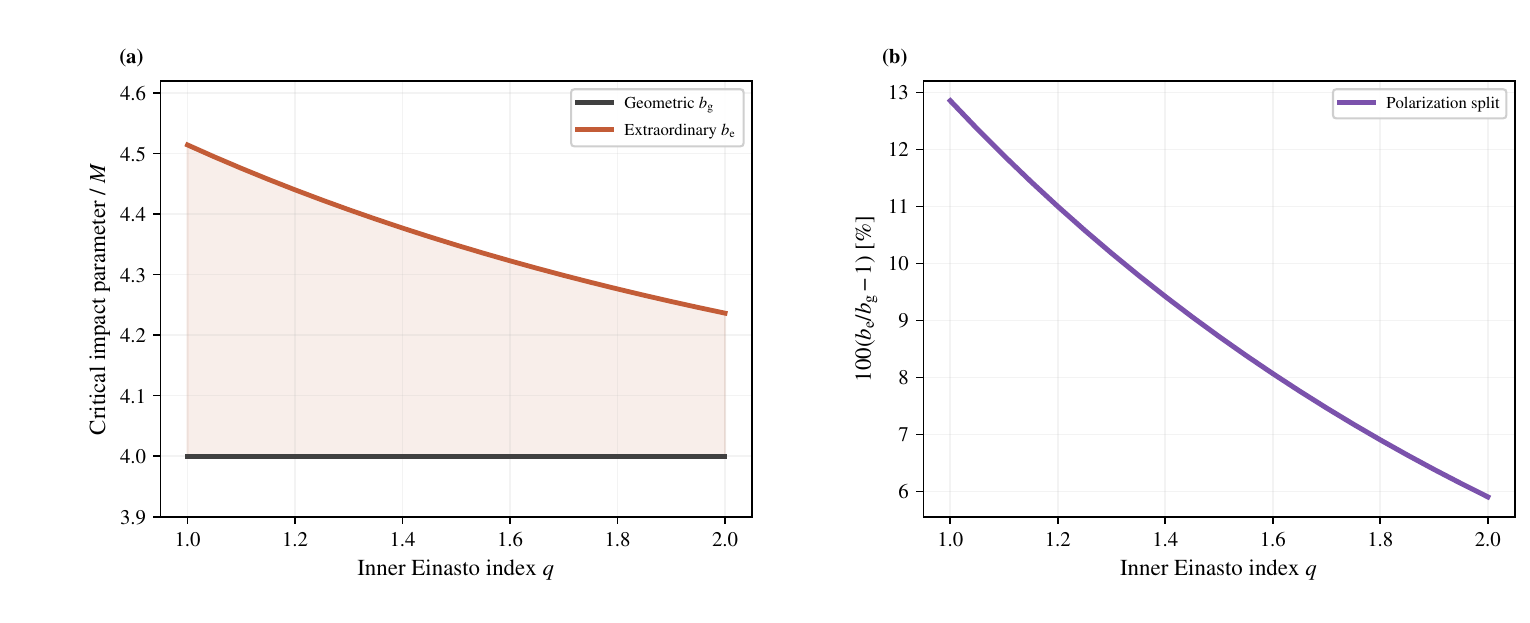}
\caption{Twenty-one magnetic solutions with the same geometric critical impact parameter. The extraordinary mode retains a distinct, $q$-dependent critical parameter despite the geometric match. Each point corresponds to a separately reconstructed NLED action.}
\label{fig:matched}
\end{figure}

\subsection{Critical points as directly checkable root conditions}
For the background cone, differentiating $U_{\rm g}=f/r^2$ gives the standard circular-orbit condition $r f'-2f=0$. Because $f=1-2m/r$ and $m'=4\pi r^2\rho$, it can also be written as
\begin{equation}\label{eq:photon-einstein}
 3\frac{m(r_{\rm g})}{r_{\rm g}}-m'(r_{\rm g})=1.
\end{equation}
This form is useful because it ties the optical radius directly to the enclosed mass and local density rather than only to a metric function. The extraordinary branch has an additional constitutive term,
\begin{equation}\label{eq:photon-extra}
 \frac{f'}{f}+\frac{\kappa'}{\kappa}-\frac2r=0,
\end{equation}
and this equation is solved only where $f>0$ and $\kappa>0$. The difference between Eqs.~(\ref{eq:photon-einstein}) and (\ref{eq:photon-extra}) is the origin of the optical splitting. Merely changing $Q$ in an otherwise Maxwellian geometry would not reproduce that second derivative term.

We check each optical root in three ways. First, a numerical root is bracketed outside the largest positive zero of $f$. Second, the computed potential has negative second derivative at that root, so it is a maximum of the relevant radial potential and a candidate capture threshold. Third, the corresponding $b_j=U_j(r_j)^{-1/2}$ is checked against direct rays with impact parameters on either side of the root. We apply these controls to both characteristics. The procedure guards against using a stable orbit or an interior root as an apparent shadow boundary. Standard shadow reviews discuss the general separatrix interpretation and the differences between the geometric and imaging definitions \cite{Perlick2022,Gralla2019}.

The matched-geometric exercise in the first part of our optical study is intentionally restricted. The $21$ solutions have $q$ values between one and two and separately adjusted $R$ values that set $b_{\rm g}=4M$. Because the reconstructed action depends on both $q$ and the normalization scale, this ensemble is not the evolution of a single black hole under a varying physical charge. It answers the question ``how much does the extraordinary characteristic change if the geometric capture scale is held fixed across related actions?'' The later fixed-action family answers the different question ``how do both capture scales change when the magnetic charge labels solutions of the same theory?'' Reporting both, with their assumptions, is more informative than treating one of them as a universal observational degeneracy.

\subsection{What the effective cone does and does not predict}
A characteristic orbit describes high-frequency propagation in this field theory. In a magnetized source the extraordinary cone describes one polarization of a high-frequency electromagnetic disturbance, whereas a gravitational perturbation has additional coupled field degrees of freedom. Consequently the extraordinary critical impact is not necessarily a second bright annulus in a synchrotron image. A realistic image depends on where radiation is generated, the local emitted polarization, redshift, Faraday transport and the interferometric response. Prior effective-metric studies have already found important differences between metric shadows and photon-effective-geometry images \cite{Walia2024,Uniyal2024,DePaula2026,GuzmanMontielBreton2024}. We keep the term ``critical impact parameter'' for our field-theory prediction and reserve ``observed ring'' for an observable constructed from an emission model.

A second distinction concerns experimental identifiability. Although a polarization-dependent characteristic splitting can be expressed as a few percent of $b_{\rm g}$, there is no general rule that the same percentage difference survives unknown angular scale, emitting plasma thickness and calibration parameters in interferometric data. The exercise below retains a freely adjustable ring diameter precisely because a theoretical fractional difference should not be compared directly with the uncertainty assigned to a fitted emission-ring diameter. Such nuisance-parameter degeneracies are a major issue in prospective photon-ring interferometry \cite{Johnson2020,GrallaLupsascaMarrone2020,Broderick2022}.

\subsection{Integrated rays, turning points and strong deflection}
We also integrate the scattering trajectories. The conformal representative of the extraordinary effective metric has $g_{tt}=-f$, $g_{rr}=1/f$, and $g_{\varphi\varphi}=r^2/\kappa$ on the equatorial plane. For either ray branch let $K_{\gmode}=1$ or $K_{\ee}=\kappa$. Its turning point is the outer root of $b^2 U_j(r_0)=1$, and the exact asymptotic bending angle in this effective-geometry approximation is
\begin{equation}\label{eq:lens}
 \widehat\alpha_j(b)=2\int_{r_0}^{\infty}\frac{bK_j(r)\,dr}{r^2\sqrt{1-b^2U_j(r)}}-\pi.
\end{equation}
The integral is evaluated after the nonsingular change $r=r_0+t^2$; we also retain an asymptotic tail correction. These integrations trace the field characteristics. They do not include emission by plasma. The optical model assumes that the two eigenpolarizations can be meaningfully separated and that geometric-optics propagation applies.

\begin{figure}[tbp]
\centering\includegraphics[width=\columnwidth]{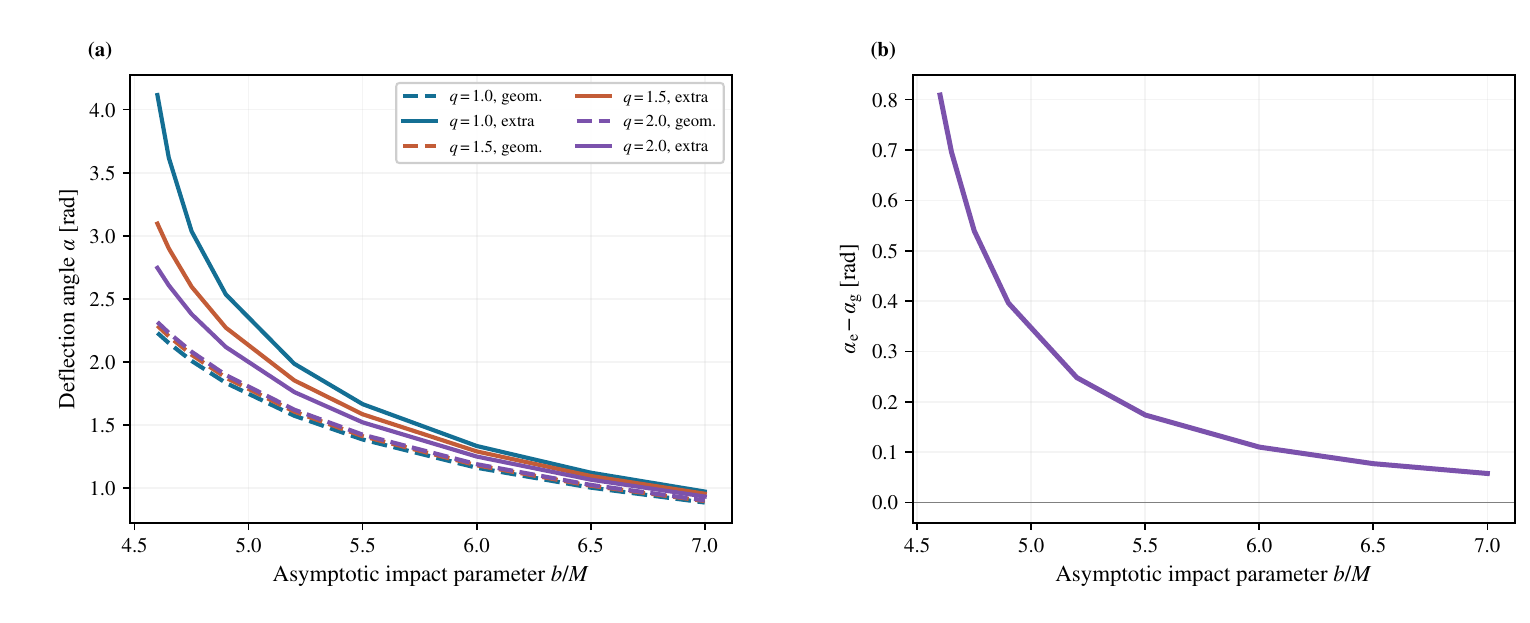}
\caption{Direct strong-deflection integrals for both polarizations in three matched-background examples (left). The right panel shows their bending-angle difference at fixed $b$ for $q=1.5$. All impact parameters are above both critical values.}
\label{fig:bend}
\end{figure}

For the middle example (Fig.~\ref{fig:bend}), $q=1.5$, the bending angles at $b=4.6M$ are $2.287$ rad (geometric) and $3.096$ rad (extraordinary), a difference of $0.809$ rad. At $b=4.9M$ the corresponding values are $1.876$ and $2.272$ rad. Such differences survive a match of the background geometric photon ring because the source changes the propagation law of one polarization.

Near the critical orbit, with $\epsilon=b/b_j-1\downarrow0$, Eq.~(\ref{eq:lens}) gives
\begin{equation}\label{eq:strong}
 \widehat\alpha_j=-\bar a_j\ln\epsilon+O(1),\qquad
 \bar a_j=\frac{\sqrt2\,K_j(r_j)}{r_j^2\sqrt{-U_j''(r_j)}}.
\end{equation}
The logarithmic limit follows the strong-deflection approach \cite{Bozza2002,BozzaScarpetta2007}. This coefficient reduces to one in the Schwarzschild limit and includes the local NLED angular factor for the extraordinary branch. For $q=1.5$, analytic values are $\bar a_{\gmode}=1.3021$ and $\bar a_{\ee}=1.2043$. A finite-range numerical fit over $\epsilon=0.002$--$0.012$ gives approximately $1.2870$ and $1.1896$; the finite-$\epsilon$ shift is consistent with the expected subleading terms. We do not claim that the finite-range fit determines the exact asymptotic limit.

\begin{figure}[tbp]
\centering\includegraphics[width=\columnwidth]{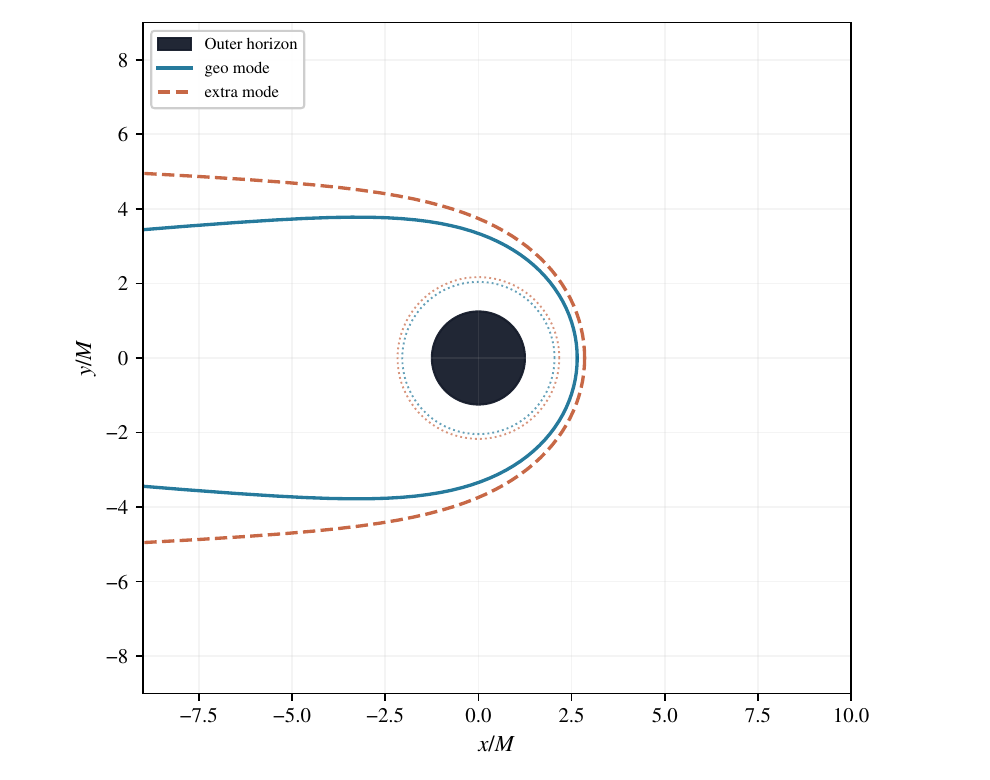}
\caption{Actual integrated equatorial characteristic rays in the matched $q=1.5$ background. Each branch uses $b=1.06b_j$ relative to its own critical curve. The dark disk marks the event horizon; dotted circles mark the respective unstable radii. These are ray trajectories, not synthetic intensity images.}
\label{fig:traj}
\end{figure}

\subsection{A check from integrated rays}
The bending integral tests the capture threshold obtained from the potential maximum. We first solve $U_j'(r_j)=0$ and calculate $b_j$. We then integrate rays with impact parameters just above this value (Fig.~\ref{fig:traj}). Their deflection increases logarithmically as $b$ approaches $b_j$. Both methods agree on the critical scale within the tested numerical accuracy. The two bending angles quoted above show that the extraordinary field can respond differently even when the geometric capture scale is fixed.

\subsection{Light-ring instability and the limits of eikonal inference}
At high multipole order, characteristic orbits can help explain wave behavior. The full coupled equations are still needed at finite frequency. This distinction is developed in analyses of general two-field eikonal systems \cite{GlampedakisSilva2019}.
The light-ring problem also has a gravitational characteristic interpretation at very high frequency. For a geometric null orbit define $\Omega_{\gmode}=\sqrt{U_{\gmode}(r_{\gmode})}=1/b_{\gmode}$ and
\begin{equation}\label{eq:lambda}
 \Lambda_{\gmode}=f(r_{\gmode})\sqrt{-\frac{U_{\gmode}''(r_{\gmode})}{2U_{\gmode}(r_{\gmode})}}.
\end{equation}
The high-frequency correspondence discussed in Ref.~\cite{Cardoso2009,Decanini2011} motivates the formal geometric estimate $\omega_{\ell n}^{(\gmode)}\simeq (\ell+1/2)\Omega_{\gmode}-i(n+1/2)\Lambda_{\gmode}$, provided the relevant high-multipole wave branch has the geometric principal symbol and is not altered by mode mixing. In our 21 models, $M\Omega_{\gmode}=0.25$ exactly by construction, but $M\Lambda_{\gmode}$ decreases from $0.20117$ at $q=1$ to $0.18609$ at $q=2$, a $7.5\%$ shift. Thus even two \emph{geometric} high-frequency scales are not fixed by a matched critical impact parameter.  Eq.~(\ref{eq:lambda}) is a photon-orbit instability diagnostic; it is \emph{not} a computed coupled Einstein--NLED gravitational quasinormal frequency.

\section{One fixed action and its physical solution family}\label{sec:fixed}

\subsection{Why parameter independence changes the question}
Once the matter action is fixed, its metric scale, charge and density normalization are linked. They cannot be chosen independently. In our case the far-field Maxwell coefficient and the central normalization are tied to the same function $\mathcal L(\mathcal F)$. To compare solutions rather than theories, we therefore fix the constitutive constants $q$ and $\rho_0$ and vary the monopole charge. The resulting mass and radius changes are consequences of the field equations. The matched-shadow family discussed above is retained as a separate control but is not used for the central charge-matched scattering inference. This distinction parallels the need to keep the physical source and its allowed parameter variations fixed when interpreting regular-black-hole solutions \cite{Bokulic2024,Bronnikov2001}.

The relation follows directly from the mass and charge equations: $M=(4\pi\rho_0R^3/q)B_q$ and $Q^2=8\pi\rho_0R^4$. Holding $\rho_0$ and $q$ fixed gives $R\propto |Q|^{1/2}$ and $M\propto R^3\propto|Q|^{3/2}$. In units of each solution's own ADM mass, this implies $R/M\propto |Q|^{-1}$ and $Q/M\propto |Q|^{-1/2}$, where the unscaled charge is measured relative to the reference member. The equations do not predict that the physical mass of an astrophysical black hole must change together with its charge; they describe a mathematical family of static solutions of the same chosen action.

The dimensionless parameter $s=|Q|/|Q_\star|$ is especially convenient because it displays which quantities are compared at fixed theory. A point at $s=1.50$ has a different ADM mass from a point at $s=1.08$ before each is nondimensionalized. For the scattering problem we use $M\omega$, $r/M$ and $Q/M$ separately for each member. Using one reference mass to nondimensionalize the whole sequence would shift the reported frequency crossings spuriously. This choice also fixes the RN comparator: its mass and charge agree with the corresponding asymptotic $1/r$ and $1/r^2$ coefficients, rather than being adjusted afterward to minimize a residual.

\subsection{Horizon, charge and constitutive admissibility of the sequence}
The branch restriction on the RN control deserves explicit attention. A magnetically sourced NLED black hole may possess a horizon even when its asymptotic charge-to-mass ratio is greater than unity; an RN solution with the same ratio does not. Such a case is a valid NLED solution but is not a valid RN \emph{black hole} comparison. Therefore the broader $27$-state exterior survey is useful for admissibility and optical trends, while the main RN-subtracted wave analysis is restricted to the eight states with $Q/M<1$. This selection prevents us from comparing black-hole absorption with an RN naked singularity or from falsely suggesting that RN exists over the entire NLED branch.

Our exterior checks are also reported at the appropriate logical level. Bracketed roots locate an inner and an outer zero of $f$ in the sampled solutions. The condition $r_h>r_\kappa$ ensures positivity of the extraordinary angular factor throughout the exterior for the particular monotone form of $\kappa$. The wave matrices are checked for real symmetry and regularity on their numerical integration grids. Positivity of finite-grid potential eigenvalues offers evidence against a simple negative barrier instability, but by itself cannot exclude all global modes or the separate known center instability of some regular magnetic configurations \cite{Moreno2003,NomuraYoshidaSoda2020,DeFelice2025}. These tests support the restricted exterior scattering calculation that is actually performed.

The matched-$b_{\gmode}$ models above vary both $q$ and $R$, and consequently vary the underlying NLED action. A more stringent physical comparison must hold the \emph{function} $\LL(\F)$ fixed. We take $q=3/2$ and choose the $b_{\gmode}=4M$ solution as a reference. For that solution $R_\star/M_\star=0.3200519151$, $Q_\star^2/M_\star^2=1.066840\ldots$, and $\F_0=4\pi\rho_0$ in the normalization of Eq.~(\ref{eq:lagrangian}). The two independent constants of the fixed action are $\rho_0$ and $q$; $\F_0=4\pi\rho_0$ is then fixed by the Maxwell limit.

For an arbitrary magnetic charge $Q$ define $s=|Q|/|Q_\star|$. The relation $\F(r)=Q^2/(2r^4)$ fixes the profile scale, while integration of the density fixes the total ADM mass. These relations are exact:
\begin{equation}\label{eq:fixed-action}
\boxed{R(s)=R_\star s^{1/2},\qquad M(s)=M_\star s^{3/2}.}
\end{equation}
For example, $R/M=(R_\star/M_\star)s^{-1}$. Within this fixed-action family, the dimensional frequency changes as $\omega=(M\omega)/M(s)\propto s^{-3/2}$ at fixed dimensionless $M\omega$. Thus a trend plotted against $M\omega$ need not have the same percentage change in physical frequency. Charge, mass and source size are \emph{not} independent fitting parameters in one magnetic theory. Equation~(\ref{eq:fixed-action}) follows from $\rho_0=qM/(4\pi R^3 B_q)$ and $Q^2=2qMR/B_q$ before setting $M=1$. Independent checks of $\rho_0$, $\F_0$, and the scaling identity were performed at each charge.

\begin{figure}[tbp]
\centering\includegraphics[width=\columnwidth]{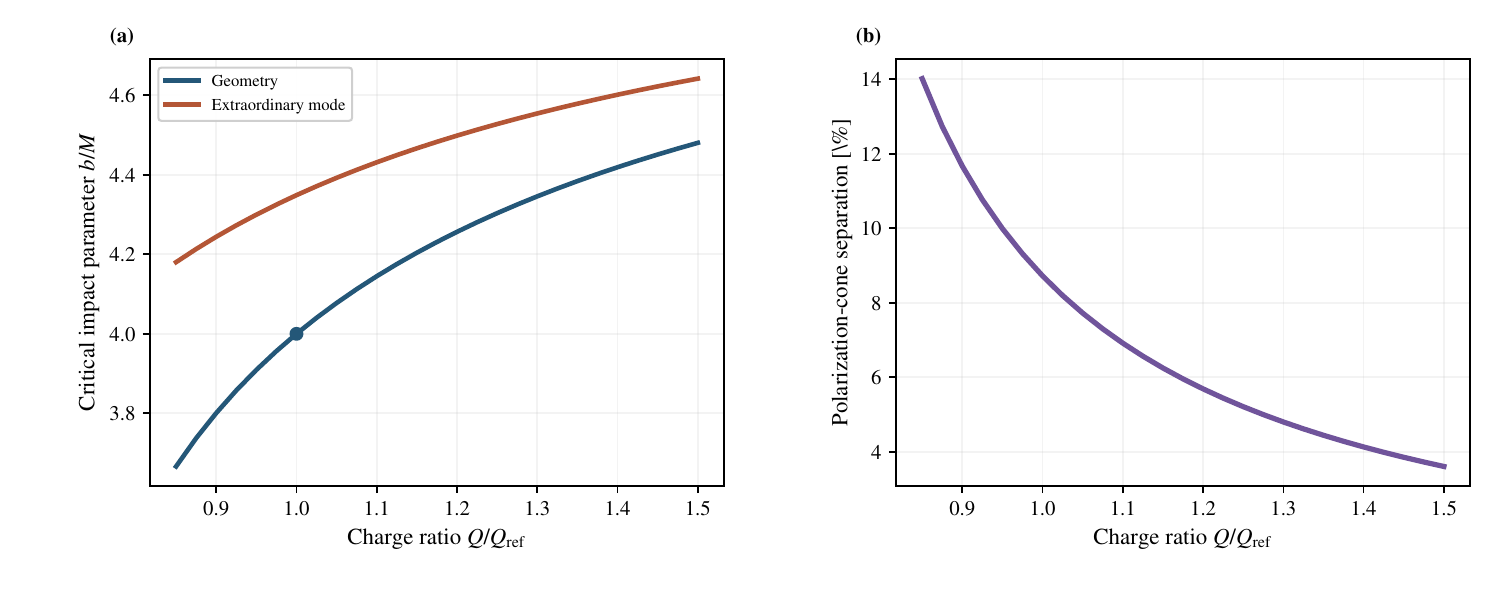}
\caption{A physically restricted magnetic family with a single fixed NLED action and $q=1.5$. (a) Both geometric and extraordinary critical scales vary as charge is changed. (b) Their fractional separation also varies. No artificial geometric-size matching is imposed.}
\label{fig:fixed-action}
\end{figure}

We evaluate 27 solutions with $0.85\le s\le1.50$ using the exact mass function and bracketed numerical orbit roots. Every sampled case has two metric horizons and positive $\kappa$ throughout its sampled exterior. The geometric impact parameter $b_{\gmode}/M$ now varies monotonically from $3.6664$ to $4.4803$ (Fig.~\ref{fig:fixed-action}); the extraordinary-to-geometric split varies from $14.01\%$ to $3.60\%$. Near the reference solution the logarithmic response is $d\ln (b_{\gmode}/M)/d\ln|Q|\simeq0.419$ (finite difference). This \emph{fixed-theory} trend is distinct from the earlier same-$b_{\gmode}$ comparison across different actions. It removes the misleading interpretation of the latter as a continuous parameter degeneracy inside one matter theory.

\section{Coupled gravitational--electromagnetic perturbations}\label{sec:coupled}
\subsection{Why the matter perturbation cannot be omitted}
The coupled perturbation problem follows the gauge-invariant Einstein--Maxwell reductions developed for charged backgrounds \cite{Moncrief1974Even,Moncrief1975,Zerilli1974RN}, with constitutive corrections from the magnetic nonlinear source \cite{Moreno2003,Pradhan2026}. The Maxwell weak-field limit gives both photon characteristics a shared asymptotic normalization. On its own, a critical null orbit is not a prediction of a finite-multipole gravitational waveform; for that we calculate the coupled field equations in the following sections. An important distinction remains: the 21 fixed-shadow optical backgrounds in Fig.~\ref{fig:matched} compare different actions, whereas the eight-state optical--wave series below keeps \emph{one} action fixed. The exact exponential Einasto density has been replaced by an electromagnetic model with the same leading inner expansion and an $r^{-4}$ tail; this model is not a collisionless dark matter halo.

The observed emission ring is not necessarily the NLED extraordinary critical curve. A polarized emissivity and propagation model is needed before assigning any observational weight to the two characteristic scales. A simple visibility illustration later in this paper is not an EHT measurement, and no inference from detector gravitational-wave strain is attempted.

\subsection{Parity pairing and degrees of freedom}
A two-component wave equation is not obtained by taking the Regge--Wheeler potential and adding an electromagnetic barrier by hand. The background carries a magnetic field, and variation of the Einstein--NLED action produces cross terms between the metric and electromagnetic perturbations. The physical combinations may be organized into parity sectors, but the magnetic field changes which electromagnetic parity is paired with a given metric parity. We therefore refer throughout to \emph{metric parity} and specify the magnetic pairing, rather than using the ambiguous phrase ``axial electromagnetic mode'' without a master-field convention.

In each sector and for $\ell\ge2$, the reduced physical system can be written using two canonical amplitudes. These amplitudes represent coupled fluctuations; unless a particular asymptotic basis is chosen, neither is an independent pure gravitational or pure electromagnetic outgoing wave at every radius. The $2\times2$ potential matrix is real and symmetric after the flux normalization used here. Its off-diagonal entries are physical mixing terms, not optional numerical corrections. A coupled perturbation system of this type has a well-established derivation in the gauge-invariant self-gravitating NLED framework \cite{Moreno2003,Chaverra2016,Pradhan2026}. We specialize published canonical expressions to our matter action, and the appendix provides the explicit even-sector entries so that the source dependence is inspectable.

The principal distinction from a test electromagnetic field on a fixed background is that the latter does not allow the electromagnetic disturbance to perturb the Einstein tensor. A test scalar or a test electromagnetic quasinormal mode can be useful as a probe of the background geometry, but it cannot replace the present gravitoelectromagnetic flux matrix. Conversely, the real-frequency transmission calculation we carry out is not necessarily a computation of complex quasinormal poles. Work on test-field optical and greybody phenomena provides context, but the dynamical quantity and boundary conditions are different \cite{Toshmatov2018,Okyay2022,Liang2026}.

\subsection{Constitutive gradients and their appearance in the wave operator}
The NLED-specific derivative terms can be expressed through a compact radial gradient rather than differentiating the action numerically at every integration point. Define $A=d\ln\mathscr L_{\mathcal F}/dr$. For the fixed profile, Eq.~(\ref{eq:LF-full}) immediately gives $A=(q+4)/[r(1+x)]$. The derivative $A'$ follows analytically, and the two second-derivative combinations $\mathcal D_\pm$ differ by the sign of $f'A+fA'$. These terms disappear in Maxwell theory, where $\mathscr L_{\mathcal F}$ is constant. Their presence in the diagonal matrix entries and the factors of $\sqrt{\mathscr L_{\mathcal F}}$ in the off-diagonal entries show explicitly which part of the wave problem is sensitive to the nonlinear constitutive law.

For a numerical implementation, one should evaluate $m(r)$ from the incomplete-beta expression, use $m'=4\pi r^2\rho$ for the first mass derivative, and compute $f'=2m/r^2-2m'/r$ without noisy finite differences. The logarithmic derivative $A$ and the optical coefficient $\kappa$ can be evaluated in closed form. The remaining matrix entries are rational combinations of $m,f,\kappa,A,A'$ and the angular eigenvalue $L=\ell(\ell+1)$. This is not just a programming convenience: differentiating a numerically tabulated mass or Lagrangian can introduce artificial oscillations in a potential, which in turn can contaminate scattering phases and spurious zero crossings.

The Maxwell limit provides a check on these coefficients. In the Maxwell limit $A\to0$, $\kappa\to1$, and $\mathscr L_{\mathcal F}\to1$. The canonical operator must then reduce to the charged RN master-field system when evaluated on an RN metric. For vanishing charge the off-diagonal entries vanish, leaving the Regge--Wheeler and Zerilli gravitational potentials and the familiar electromagnetic multipole barrier. These limits constrain normalization, parity signs and the angular factor $\lambda=(\ell-1)(\ell+2)$. We checked these limits algebraically using the same source coefficients as in the numerical operator. The checks test this operator, without rederiving the complete gauge-invariant reduction.

\subsection{Canonical parity sectors and potential matrices}
In the coupled problem, the magnetic field and metric both fluctuate. Test-field analyses of nonlinear-electromagnetic black holes and their greybody factors provide useful but distinct reference calculations \cite{Toshmatov2018,Okyay2022}. We specialize the canonical Einstein--nonlinear-electrodynamics (NLED) perturbation matrices of Moreno and Sarbach \cite{Moreno2003} and Pradhan et al. \cite{Pradhan2026} to the single-action magnetic family derived above. This specialization is not a new derivation of the general field equations. We use $\mathscr L=4\pi\rho$, $\mathscr L_{\mathcal F}=4\pi\LL_{\mathcal F}$, $L=\ell(\ell+1)$, and $\lambda=L-2$. The reduced flux-normalized system in each metric-parity sector $\mathcal P=\pm$ is
\begin{equation}
\left[-\partial_t^2+\partial_{r_*}^2-\mathbf V_{\mathcal P}(r)\right]\mathbf\Psi_{\mathcal P}=0,\quad dr_*=dr/f,
\end{equation}
where $\mathbf V_{\mathcal P}=f\mathbf U_{\mathcal P}$ is symmetric and acts on gravitational and electromagnetic amplitudes. For a magnetic background, odd metric perturbations couple to even electromagnetic perturbations, and even metric perturbations to odd electromagnetic perturbations; the parity labels on $\mathbf V_{\mathcal P}$ refer to the metric. The odd metric-parity matrix is
\begin{align}
U_-^{gg}&=L/r^2-6m/r^3+2\mathscr L,\\
U_-^{ge}&=2Q\sqrt{\lambda\mathscr L_{\mathcal F}}/r^3,\\
U_-^{ee}&=L/r^2+4Q^2\mathscr L_{\mathcal F}/r^4+\mathcal D_-,
\end{align}
where $\mathcal D_-=\mathscr L_{\mathcal F}^{1/2}\partial_r[f\partial_r\mathscr L_{\mathcal F}^{-1/2}]$. The even metric-parity matrix is given explicitly in Appendix~\ref{app:even}. Our fixed action has a useful closed-form constitutive response:
\begin{align}
\mathscr L_{\mathcal F}&=\left(\frac{x}{1+x}\right)^{1+4/q},\quad A\equiv\partial_r\ln\mathscr L_{\mathcal F}=\frac{q+4}{r(1+x)},\\
\mathcal D_\pm&=\tfrac14f A^2\pm\tfrac12(f'A+fA'),\\
A'&=-\frac{(q+4)[1+(q+1)x]}{r^2(1+x)^2},\quad x=(r/R)^q.
\end{align}
The plus and minus derivatives refer to the respective metric parity. In the Einstein--Maxwell limit, $A=0$ and $\kappa=1$; with the independent RN background the matrices reproduce the standard charged-wave problem. The $Q=0$ RN limit recovers uncoupled Schwarzschild gravitational and electromagnetic potentials. These are analytic reference controls, not an extrapolation along the fixed-action $M\propto Q^{3/2}$ family.

\section{Flux-normalized real-frequency scattering}\label{sec:scattering}
We first check $\ell=2$ on the 27 fixed-action states of the previous section. All sampled exterior configurations satisfy $\mathscr L_{\mathcal F}>0$, $\kappa>0$, and $0<f\kappa<3$, with positive eigenvalues of both potential matrices on a dense exterior grid. The smallest sampled dimensionless potential eigenvalue is $4.33\times10^{-6}$. The exterior conditions are consistent with known sufficient criteria \cite{Moreno2003}; finite-grid positivity is not an all-radius proof. More importantly, these exterior checks must not be conflated with global stability: instability around the regular center in single-invariant NLED has been demonstrated elsewhere \cite{DeFelice2025}.

For each real positive frequency we integrate two independent ingoing horizon solutions simultaneously and match their amplitudes at a distant radius to incoming and outgoing plane waves. The reflection and transmission matrices are
\begin{equation}
\mathbf R=\mathbf A_{\rm out}\mathbf A_{\rm in}^{-1},\quad \mathbf T=\mathbf A_{\rm in}^{-1},\quad \mathbf R^\dagger\mathbf R+\mathbf T^\dagger\mathbf T=\mathbf I.
\end{equation}
We diagonalize $\mathbf T^\dagger\mathbf T$ and report the smaller and larger absorption eigenvalues $\Gamma_{\rm low}$ and $\Gamma_{\rm high}$. Unlike individual off-diagonal conversion coefficients, these eigenvalues are invariant under unitary changes of the asymptotic incoming channel basis. The maximum flux-conservation residual over our reference scans is $5.0\times10^{-7}$.

At $M\omega=0.45$, the $s=1.2$ black hole gives $\Gamma_{\rm high}^{-}=0.7185$ and $\Gamma_{\rm high}^{+}=0.7671$. Their difference, $\Delta\Gamma=0.0485$, is a direct finite-frequency coupled-wave result. The corresponding differences for $s=1.0$ and $1.5$ are $0.1072$ and $0.0205$, respectively. The frequency dependence is studied below. These representative values are collected in Table~\ref{tab:absorption-snapshot}.
\begin{table}[tbp]
\caption{Larger-eigenvalue absorption at $M\omega=0.45$ in a single fixed-action family; RN control black holes are available only when $Q/M<1$.}\label{tab:absorption-snapshot}
\centering\small
\begin{tabular}{@{}rrrrr@{}}\toprule
$s$&$Q/M$&$\Gamma^-_{\rm high}$&$\Gamma^+_{\rm high}$&$\Delta\Gamma$\\\midrule
1.0&1.033&0.5794&0.6866&0.1072\\
1.2&0.943&0.7185&0.7671&0.0485\\
1.5&0.843&0.8025&0.8230&0.0205\\
\bottomrule\end{tabular}
\end{table}
To isolate source-dependent parity breaking, we repeat the calculation for Reissner--Nordstr\"om (RN) black holes of the same asymptotic $Q/M$ \cite{Moncrief1974,OuldElHadj2022}. RN is an admissible black-hole comparator for $s=1.2$ and $1.5$, but not for $s=1.0$, whose inferred RN charge would exceed the extremality bound. In the Schwarzschild and RN benchmarks, the two transmission-eigenvalue spectra agree between parities within finite-radius numerical error (below $2.3\times10^{-5}$ in the three benchmark frequencies). In contrast, the magnetic NLED differences persist at the percent level in the displayed wave band.

\subsection{Horizon normalization, Jost matrices and conserved flux}
It is useful to state what the scattering code solves, since numerical absorption probabilities alone do not reveal their origin. With a real frequency $\omega>0$, write $\mathbf\Psi(t,r)=e^{-i\omega t}\mathbf y(r)$. The canonical system becomes $d^2\mathbf y/dr_*^2+[\omega^2\mathbf I-\mathbf V(r)]\mathbf y=0$. At a nonextremal outer horizon $r_h$, $f$ has a simple zero and $r_*\to-\infty$. The physical black-hole boundary condition selects purely ingoing waves proportional to $e^{-i\omega r_*}$ there. A convenient fundamental solution matrix has two linearly independent ingoing horizon columns; its asymptotic expansion at large positive $r_*$ defines incoming and outgoing amplitude matrices.

A two-channel calculation must keep track of the incident mixture of gravitational and electromagnetic disturbances. In this basis, one absorption eigenvalue describes the most strongly absorbed incident combination and the other the least absorbed one; the eigenvectors specify those combinations. The eigenvalues describe energy flux, not a probability for observing a particular polarization on the sky.

The use of a fundamental matrix matters for channel mixing. If the horizon data are combined into a single arbitrarily chosen amplitude vector, one recovers only one column of the scattering transformation and cannot form its basis-invariant absorption eigenvalues. With two independent horizon columns, the large-radius matrix can be written schematically as
\begin{equation}\label{eq:jost-expansion}
 \mathbf Y(r_*)=e^{-i\omega r_*}\mathbf A_{\rm in}
                  +e^{+i\omega r_*}\mathbf A_{\rm out}+o(1).
\end{equation}
The incoming basis is invertible at generic real frequencies. Right multiplication by $\mathbf A_{\rm in}^{-1}$ produces unit incoming amplitudes, and gives the reflection and transmission matrices used in the text. The asymptotic field basis has to be flux-normalized; otherwise an apparent ``absorption probability'' can be changed by rescaling one component.

For a real symmetric potential, the matrix Wronskian is conserved. One convenient current is $\mathbf J=(\mathbf Y^\dagger\partial_{r_*}\mathbf Y-\partial_{r_*}\mathbf Y^\dagger\mathbf Y)/(2i)$, with the sign chosen consistently for incident and transmitted directions. Substituting the plane-wave expansions yields the identity $\mathbf R^\dagger\mathbf R+\mathbf T^\dagger\mathbf T=\mathbf I$ in the normalized convention. This identity is necessary but not sufficient for a correct solution: it can hold for a wrong potential if that potential is real and symmetric. We therefore combine it with analytic limiting models, boundary-radius tests and a separate solver implementation. The conversion of electromagnetic and gravitational waves by RN black holes provides a pertinent charged-wave benchmark \cite{Moncrief1974,OuldElHadj2022}.

\subsection{What the absorption eigenvalues measure}
Define $\mathbf H=\mathbf T^\dagger\mathbf T$. Flux conservation and positivity imply $0\le\Gamma_1\le\Gamma_2\le1$ for its two real eigenvalues. The larger eigenvalue is the greatest absorption possible for a normalized coherent combination of the two incoming fields. It can differ from the absorption of a purely gravitational or electromagnetic input. The maximizing combination can also change with frequency and charge.

A unitary change of the incoming basis transforms $\mathbf H$ to $\mathbf U^\dagger\mathbf H\mathbf U$. The eigenvalues remain the same, whereas a component such as $T_{12}$ depends on the basis. We therefore compare both absorption eigenvalues across parity sectors. In the RN control their spectra agree, as expected for Einstein--Maxwell theory \cite{Chaverra2016,Nomura2022}. The trace is also basis independent. It is the sum of the absorption probabilities of the two orthogonal incident fields, or twice the absorption of an equal incoherent mixture.

We study real-frequency absorption at $\ell=2$ for eight states, and at $\ell=3$ for four states. The central comparison uses the two parity sectors of each magnetic black hole. It does not depend on an RN subtraction.

\subsection{Parity contrast without a reference black hole}\label{sec:paritycontrast}
For each parity sector, let $\Gamma_{\rm high}^{\pm}$ be the larger
of the two eigenvalues of $\mathbf T_{\pm}^{\dagger}\mathbf T_{\pm}$.
We compare the two sectors of the \emph{same} black hole:
\begin{equation}\label{eq:paritycontrast}
\Delta\Gamma_{\rm par}(s,\omega)=
\Gamma_{\rm high}^{+}(s,\omega)-\Gamma_{\rm high}^{-}(s,\omega).
\end{equation}
The difference is defined within one black hole, without a reference metric. Its value does not change under a unitary rotation of the incident-channel basis. Each eigenchannel can mix gravitational and electromagnetic fields. The two parities agree numerically in the Einstein--Maxwell control.

The lower-frequency scan covers 31 frequencies in $0.30\le M\omega\le0.60$ for all eight states. The larger-eigenvalue gap is positive at all 248 sampled points. Its largest locally refined peak is $0.0900$ at $s=1.08$ and $M\omega\simeq0.426$. RN remains parity symmetric to better than $6\times10^{-5}$ on this scan. These are differences of probabilities, not relative percentages. Figures~\ref{fig:optical_wave_fingerprint} and \ref{fig:wave-refined} display these data.

\begin{table}[tbp]
\caption{Refined $\ell=2$ parity difference in the fixed-action family.
Peaks are local maxima in $0.36\le M\omega\le0.50$, obtained by wave integrations. The means use the 31-point trapezoidal rule over $0.30\le M\omega\le0.60$. All differences are in percentage points.}
\label{tab:parity-refined}
\centering\small
\begin{tabular}{@{}rrrr@{}}\toprule
$s$ & Peak (pp) & $M\omega_{\rm peak}$ & Band mean (pp)\\\midrule
1.08 & 9.00 & 0.426 & 3.34\\
1.10 & 8.48 & 0.424 & 3.15\\
1.15 & 7.37 & 0.422 & 2.74\\
1.20 & 6.46 & 0.419 & 2.40\\
1.25 & 5.71 & 0.416 & 2.13\\
1.30 & 5.09 & 0.413 & 1.90\\
1.40 & 4.14 & 0.410 & 1.54\\
1.50 & 3.42 & 0.408 & 1.27\\
\bottomrule
\end{tabular}
\end{table}

\subsection{Frequency refinement and incident-channel absorption}\label{sec:freq-refinement}
We tested the parity difference with a finer frequency grid. We solved the coupled equations at 31 frequencies from $M\omega=0.30$ to $0.60$ for each of the eight fixed-action states. This consists of 992 wave integrations, including both parities and the charge-matched RN controls. We then optimized the parity gap within $0.36\le M\omega\le0.50$. Table~\ref{tab:parity-refined} reports the local peaks and band means. The highest local peak is $0.08997$ at $M\omega=0.4258$ for $s=1.08$. It is close to the ten-frequency sampled maximum of $0.0889$.

\begin{figure*}[tbp]
\centering
\includegraphics[width=.98\textwidth]{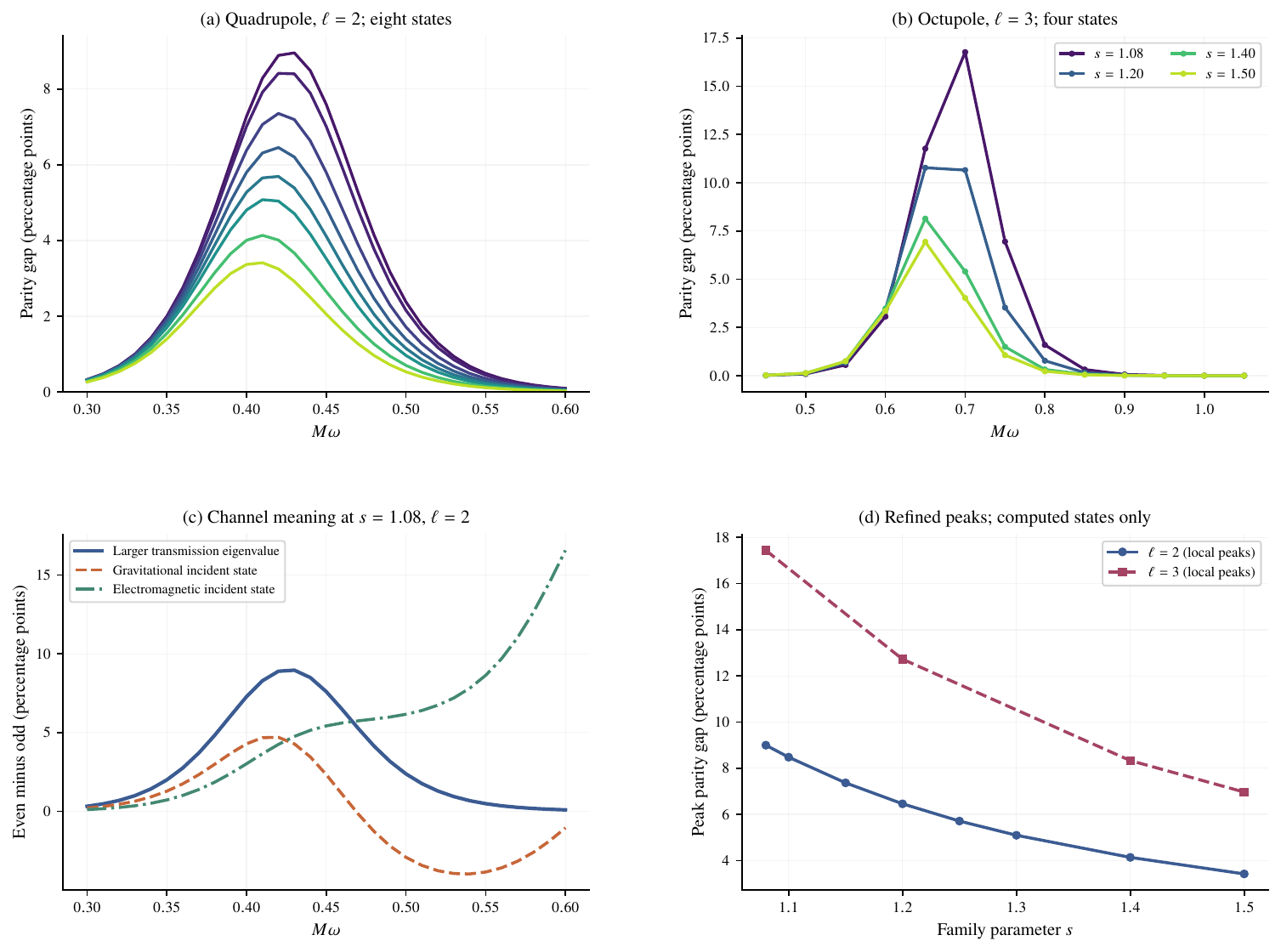}
\caption{Coupled-wave integrations and parity contrasts. (a) Direct even-minus-odd gap between the larger transmission eigenvalues, for $\ell=2$ and eight states on a 31-frequency grid. (b) The corresponding $\ell=3$ gap for four states, with 13 frequencies per state. (c) For a representative $\ell=2$ state, the larger-eigenvalue gap is compared with absorption differences for purely gravitational and purely electromagnetic incident fields in the asymptotic canonical basis. (d) Local peaks in the stated frequency intervals, $0.36$--$0.50$ ($\ell=2$) and $0.56$--$0.79$ ($\ell=3$). Lines joining the sparse $\ell=3$ points only guide the eye. All vertical axes show percentage-point differences between probabilities, not fractional percentage changes.}
\label{fig:wave-refined}
\end{figure*}

The larger eigenvalue gives the largest absorption among normalized coherent incident combinations. The diagonal entries of $\mathbf H=\mathbf T^\dagger\mathbf T$ instead give the absorption of a specified normalized incident gravitational or electromagnetic field in our asymptotic channel basis. For $s=1.08$ at $M\omega=0.42$, their parity differences are $4.70$ and $4.25$ percentage points, respectively, whereas the larger-eigenvalue difference is $8.89$ percentage points. At the upper end of the tested band, $M\omega=0.60$, the electromagnetic-input difference is $16.56$ percentage points and is still increasing there. We therefore do not call that endpoint a frequency-optimized peak. For example, at $s=1.08$ and $M\omega=0.42$, the gravitational-input probabilities are $0.3183$ (odd) and $0.3653$ (even). The electromagnetic-input probabilities are $0.0731$ and $0.1155$. Both inputs have a parity-dependent response, but their differences need not equal the gap between the larger eigenvalues. Figure~\ref{fig:wave-refined}(c) shows this distinction across the frequency band. This matters if the calculation is later connected to a physical incident-wave source.

We also made a separate $\ell=3$ test. Four states were solved at 13 frequencies between $M\omega=0.45$ and $1.05$, with RN controls. Local optimization within $0.56$--$0.79$ gives parity gaps of $0.1744$, $0.1272$, $0.0833$, and $0.0697$ for $s=1.08$, $1.20$, $1.40$, and $1.50$, respectively. The largest absolute RN parity gap in the sampled $\ell=3$ grid is below $8.4\times10^{-6}$. The maximum flux-conservation residual in that grid is below $7.2\times10^{-7}$. These results show that the parity difference is not restricted to $\ell=2$. The $\ell=3$ peak is in a different frequency range, so the two multipoles should not be compared at one arbitrarily chosen frequency. We do not infer a result for the uncomputed higher multipoles.

To check the numerical matching, we changed the outer matching radius from $240M$ to $360M$ and repeated the local peak calculations with smaller horizon offsets and tighter integration tolerances. The largest change among these tests is below $7\times10^{-5}$ in the dimensionless peak probability difference. This is much smaller than the effect being studied. The tests support the quoted two- to three-significant-digit values; they do not constitute a full analytic error bound. The wave equation and its flux normalization remain those introduced in Sec.~\ref{sec:scattering} \cite{Moreno2003,Chaverra2016,Pradhan2026}.

\subsection{The second eigenchannel and total absorption}\label{sec:trace-gap}
The larger absorption eigenvalue is not the full response. The transmission matrix has two eigenvalues in each sector. We define their separate parity differences and their sum:
\begin{align}\label{eq:tracegap}
\Delta\Gamma_{j}&=\Gamma_{j}^{+}-\Gamma_{j}^{-},\qquad j\in\{\mathrm{high},\mathrm{low}\},\\
\Delta\mathcal T&=\mathrm{Tr}\,\mathbf H_{+}-\mathrm{Tr}\,\mathbf H_{-}
=\Delta\Gamma_{\rm high}+\Delta\Gamma_{\rm low}.
\end{align}
In the asymptotic gravitational--electromagnetic basis, the same trace is $\Gamma_{gg}+\Gamma_{ee}$. This identity is an independent check on the data. It also gives a simple meaning to the trace: $\mathcal T/2$ is the absorption of an equal incoherent mixture of the two incident fields. Unlike the larger eigenvalue, the trace does not select an incident field separately at each frequency.

At $s=1.08$ and $M\omega=0.42$, the larger-eigenvalue parity gap is about $8.89$ percentage points, while the smaller-eigenvalue gap is only about $0.05$ percentage points. At $M\omega=0.60$, they become about $0.09$ and $15.42$ percentage points, respectively. Thus the parity effect moves between the two eigenchannels as the frequency increases. It has not disappeared.

We extended the $\ell=2$ wave calculation for all eight states with 20 further frequencies in $0.62\le M\omega\le1.00$. Each new point was obtained by solving the coupled radial equations in both sectors. The extended magnetic calculation adds 320 model--parity--frequency integrations. The 31-point and 20-point grids form a 51-frequency dataset for each state. The trace gap rises to a maximum and later decreases when both eigenchannels approach full absorption. Its peak is not found by extrapolating the original band.

\begin{table}[tbp]
\caption{The optical split, the locally optimized larger-eigenvalue peak in $0.36\le M\omega\le0.50$, and the trace peak in the extended $0.30\le M\omega\le1.00$ band. The absorption differences are in percentage points.}
\label{tab:R53-trace-peaks}
\centering\footnotesize\setlength{\tabcolsep}{3pt}
\begin{tabular}{@{}rrrrr@{}}\toprule
$s$ & Opt. (\%) & High (pp) & Trace (pp) & $M\omega_\mathrm{tr}$\\\midrule
1.08 & 7.22 & 9.00 & 24.35 & 0.645 \\
1.10 & 6.91 & 8.48 & 23.34 & 0.639 \\
1.15 & 6.25 & 7.37 & 21.14 & 0.627 \\
1.20 & 5.69 & 6.46 & 19.29 & 0.617 \\
1.25 & 5.21 & 5.71 & 17.73 & 0.608 \\
1.30 & 4.80 & 5.09 & 16.39 & 0.600 \\
1.40 & 4.13 & 4.14 & 14.19 & 0.586 \\
1.50 & 3.60 & 3.42 & 12.48 & 0.575 \\
\bottomrule\end{tabular}
\end{table}

The largest value in this family is about $24.3$ percentage points at $s=1.08$, near $M\omega=0.645$. The peak is smaller at larger $s$, as shown in Table~\ref{tab:R53-trace-peaks}. This is the main wave result of the extended analysis. The roughly nine-percentage-point peak of the larger eigenvalue remains a useful but incomplete part of the spectrum.

\subsection{What produces the parity gap?}\label{sec:parity-mechanism}
The potential matrices show why the two sectors may absorb differently. In the even metric-parity sector, the electromagnetic diagonal entry begins with the extraordinary angular term $\kappa\ell(\ell+1)/r^2$, while the odd sector begins with the ordinary angular term $\ell(\ell+1)/r^2$. The exterior solution has $0<\kappa<1$. The even-sector electromagnetic barrier is therefore reduced at leading angular order. The other diagonal terms and the gravitational--electromagnetic coupling also contribute.

To identify which coefficient matters most, we made a formal operator test on the \emph{same} background metric. Setting $\kappa=1$ in the wave operator, while leaving its other coefficients unchanged, reduces the trace gap at $s=1.08$, $M\omega=0.64$, from $24.22$ to about $0.05$ percentage points. Setting only the logarithmic-derivative terms $A,A'$ to zero leaves a gap of about $23.73$ percentage points. At $s=1.20$ the same test gives a similar pattern. These operations are not new physical solutions of the field equations. They are controlled changes of the operator that identify its sensitivity to the extraordinary characteristic response. In particular, the result does not follow from the geometry alone.

\begin{figure*}[tbp]
\centering\includegraphics[width=.96\textwidth]{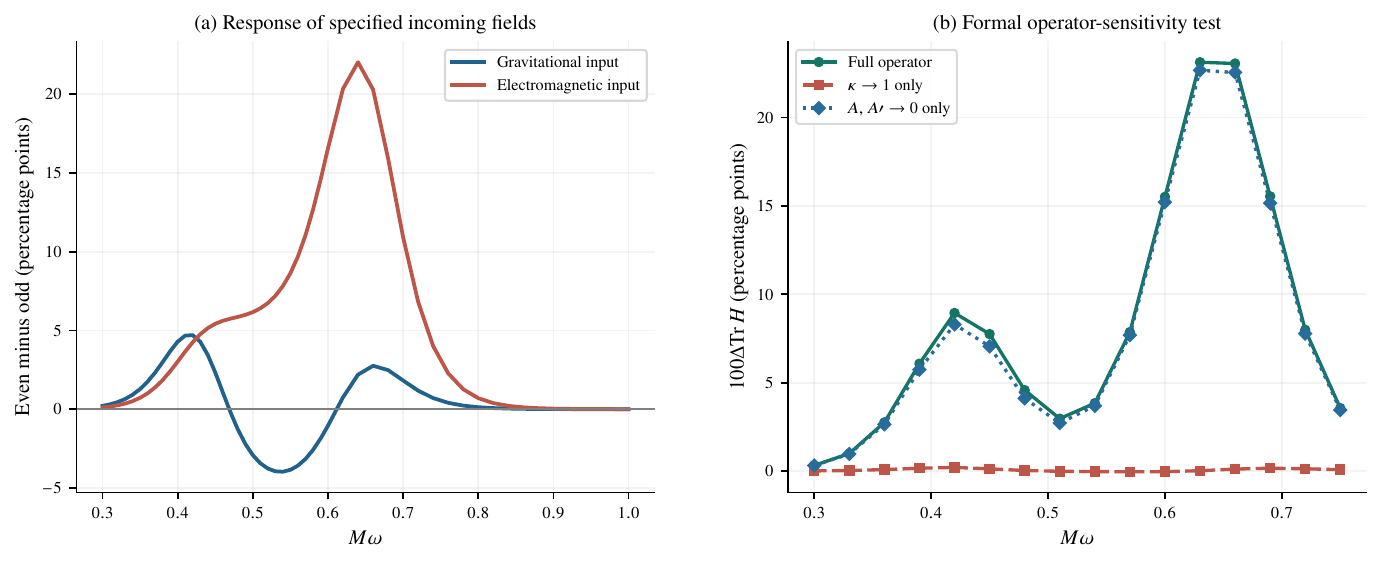}
\caption{Two views of the parity response at $s=1.08$. (a) Differences for specified gravitational and electromagnetic incident waves. Their frequency dependence differs from that of either absorption eigenvalue. (b) Formal operator tests with the background metric held fixed. Changing $\kappa$ strongly reduces the trace gap, while suppressing only $A,A'$ has a smaller effect. The modified operators are diagnostic tests, not independent black-hole models.}
\label{fig:R53-mechanism}
\end{figure*}

Figure~\ref{fig:R53-mechanism}(a) also shows that a specified gravitational input need not have the same gap as an electromagnetic input. These are physically defined asymptotic channels. The eigenvalue gaps instead describe coherent combinations chosen by the transmission operator. The two statements are compatible, but they answer different questions.

\subsection{Charge-matched RN as a secondary control}\label{sec:joint}
We also compare the magnetic black holes with RN solutions at the same ADM mass and charge. The direct parity difference remains the main wave measure. That charge is fixed by the metric and is not adjusted to fit a spectrum. RN provides a useful Einstein--Maxwell control because its two parity spectra agree in our numerical calculation \cite{Moncrief1974,OuldElHadj2022}.

For either metric-parity sector we define
\begin{equation}\label{eq:rnresidual}
\delta\Gamma_{\pm}(s,\omega)=\Gamma_{\rm high}^{\rm NLED,\pm}(s,\omega)
-\Gamma_{\rm high}^{\rm RN,\pm}(s,\omega).
\end{equation}
The largest sampled even-sector residual in the tested $\ell=2$ band is $0.1036$ at $s=1.08$ and $M\omega=0.42$. This residual includes changes in both the metric and the nonlinear constitutive response. The direct parity difference does not depend on the RN reference.

The calculation uses eight black holes and 31 frequencies for both parity sectors and both metric models. Each frequency calculation integrates two independent ingoing solutions of the coupled system. On the lower-frequency band there are $8\times31\times2\times2=992$ magnetic and formal RN-control integrations. We use RN as a black-hole comparator only for states below its charge extremality limit. The extended band adds 320 magnetic integrations. The magnetic trace dataset therefore contains 816 state--parity--frequency integrations. The ten-frequency survey and 19-point checks give additional tests of frequency sampling. Table~\ref{tab:parity-refined} reports the 31-point larger-eigenvalue results, while Table~\ref{tab:R53-trace-peaks} contains the extended trace peaks. The state-resolved ten-frequency RN residuals appear in Tables~\ref{tab:wave-residual-108}--\ref{tab:wave-residual-150}.

\subsection{Frequency structure of the RN control}
The RN comparison also varies with frequency. In each of the eight states, the odd-sector residual crosses zero within the tested $\ell=2$ band, while the even-sector residual is positive at that frequency. A zero here means equality with RN absorption for that parity sector; it does not mean that the total absorption vanishes. The roots come from a separate bracketed wave calculation, rather than from a visual interpolation of a curve.

The sign of the band-averaged odd residual depends on its positive and negative contributions. This is why the ten-point and 19-point surveys are useful checks. Our main summary uses the 31-point grid. Its mean even-sector RN residual decreases from $3.69$ to $1.22$ percentage points over the eight-state sequence, whereas the signed odd-sector mean ranges from about $+0.36$ to $-0.05$ percentage points. These statements concern the stated uniform band, not a detector-weighted average.

\section{Numerical methods and accuracy}\label{sec:validation}
We compute the background mass with the incomplete beta function and use analytic derivatives for the metric and constitutive coefficients. The wave solver integrates two independent ingoing solutions in each metric-parity sector. At large radius it extracts the incoming and outgoing amplitudes and forms the flux-normalized transmission matrix. We test the same frequencies in the magnetic model and in the charge-matched RN control.

The original $\ell=2$ grid has 31 frequencies between $M\omega=0.30$ and $0.60$ for eight states. The extended calculation adds 20 frequencies between $0.62$ and $1.00$ in each state. The two grids together determine both eigenvalues and the trace gap. The $\ell=3$ calculation still covers four states in a separate band. A separate two-frequency spot check is tabulated in Table~\ref{tab:app-l3-spot}. Local wave integrations refine the peaks found on these grids.

We checked the effect of frequency sampling before using the 31-point results. At $s=1.08$ and $s=1.50$, changing from 10 to 19 frequencies changes the even-sector RN-residual mean from $0.037555$ to $0.036926$ and from $0.012390$ to $0.012203$. These shifts are $1.7\%$ and $1.5\%$ relative to the 19-point values. The main parity means are then calculated on the 31-point grid for every state.

We also changed the numerical boundaries. For $s=1.20$ at $M\omega=0.45$, moving the outer matching radius from $150M$ to $800M$ changes the parity contrast by less than $1.2\times10^{-4}$. Reducing the starting distance from the horizon from $10^{-4}M$ to $10^{-5}M$ changes it by less than $10^{-5}$. These tests measure boundary sensitivity. Flux conservation provides a separate check of numerical normalization. Over the 320 additional magnetic integrations, the largest flux-closure residual is below $7\times10^{-7}$. For $s=1.08$ near the trace peak, moving the outer matching radius between $150M$ and $360M$ changes the trace gap by less than $4.2\times10^{-5}$. A tighter integration tolerance and a smaller horizon offset preserve the quoted peak to the shown precision.

Finally, a second Runge--Kutta implementation was used for selected solutions. It reproduced the optical splitting of $7.21697\%$ and the RN-subtracted even-sector residual $0.1035967$ at $s=1.08$ and $M\omega=0.42$. At $s=1.20$ and $M\omega=0.45$, it gave $\Gamma^-_{\rm high}=0.718526$ and $\Gamma^+_{\rm high}=0.767070$. Its largest flux residual in the checked wave cases was $3.1\times10^{-8}$. The numerical comparisons and their scope are collected in Table~\ref{tab:error-budget-r45}.

\begin{table*}[!t]
\caption{Targeted numerical checks. The values refer to the indicated test, not to a universal error bound.}\label{tab:error-budget-r45}
\centering\small
\begin{tabular}{@{}lll@{}}\toprule
\parbox[t]{.20\textwidth}{\raggedright Test\strut} & \parbox[t]{.25\textwidth}{\raggedright Result\strut} & \parbox[t]{.44\textwidth}{\raggedright Interpretation\strut} \\\midrule
\parbox[t]{.20\textwidth}{\raggedright Flux conservation\strut} & \parbox[t]{.25\textwidth}{\raggedright $\lesssim1.6\times10^{-8}$ in the coarse scan\strut} & \parbox[t]{.44\textwidth}{\raggedright Check of the scattering normalization\strut} \\
\parbox[t]{.20\textwidth}{\raggedright Boundary displacement\strut} & \parbox[t]{.25\textwidth}{\raggedright $<1.2\times10^{-4}$ in a representative case\strut} & \parbox[t]{.44\textwidth}{\raggedright Sensitivity to the outer and near-horizon matching points\strut} \\
\parbox[t]{.20\textwidth}{\raggedright Frequency integration\strut} & \parbox[t]{.25\textwidth}{\raggedright $1.7\%$, $1.5\%$ in two 10-to-19 point tests\strut} & \parbox[t]{.44\textwidth}{\raggedright Separate checks of the band mean; main table uses 31 points\strut} \\
\parbox[t]{.20\textwidth}{\raggedright Independent solver\strut} & \parbox[t]{.25\textwidth}{\raggedright $0.1035967$ RN residual reproduced\strut} & \parbox[t]{.44\textwidth}{\raggedright Check against a separate numerical code path\strut} \\
\bottomrule\end{tabular}
\end{table*}

\section{Connection to observational inference}\label{sec:observation}
Our optical results concern characteristic orbits, not bright rings measured by a telescope. Emission depends on the plasma, its magnetic field and the transfer of polarized light. In particular, the lensed emission diameter of M87* is not necessarily either optical critical scale \cite{EHT2019,EHTSgr2022,EHT2025,Gralla2019,Johnson2020,EHTpolar2021}. We use a simple visibility model to illustrate one relevant degeneracy.

\begin{figure}[tbp]
\centering\includegraphics[width=\columnwidth]{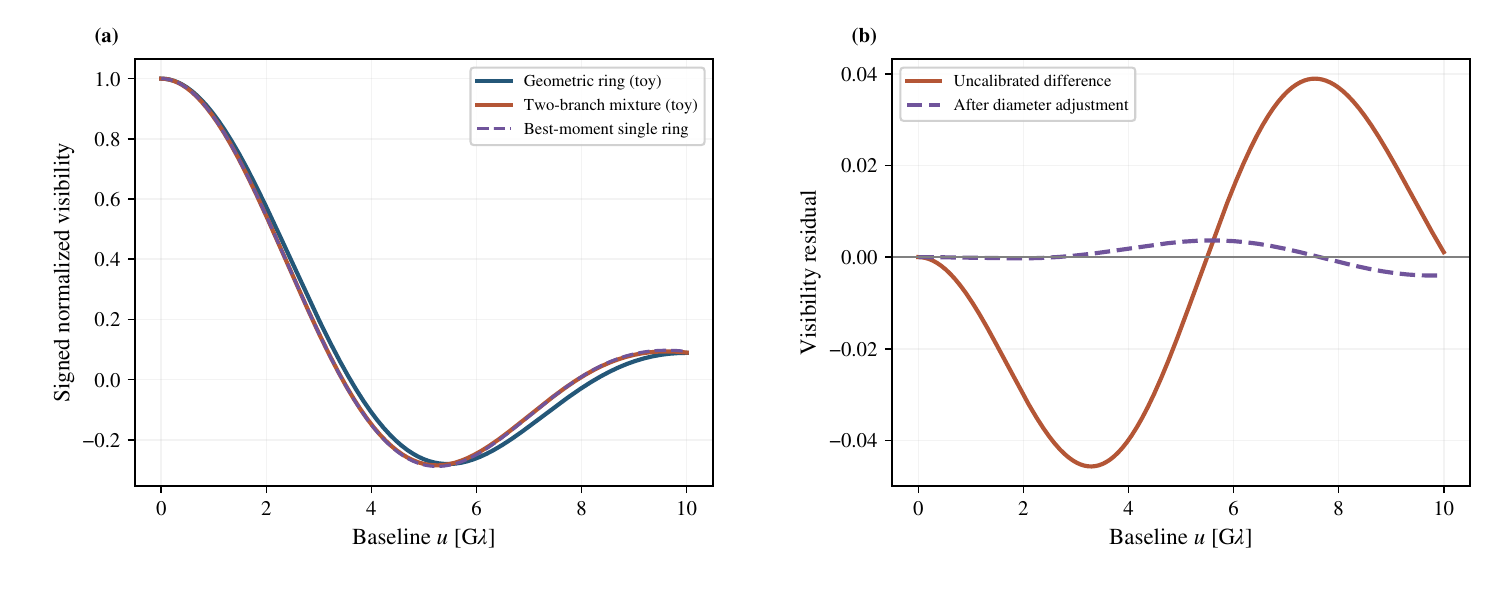}
\caption{Illustrative two-ring visibility model. The difference from a one-ring curve becomes much smaller when the mean diameter is allowed to vary. These curves are theoretical and contain no interferometric noise model.}\label{fig:vis}
\end{figure}

For a circular ring of angular diameter $D$, the normalized visibility is proportional to $J_0(\pi D u)$, with $u$ the projected baseline. For two illustrative contributions we take
\begin{equation}\label{eq:toyVis}
V(u)=e^{-2\pi^2\sigma^2 u^2}\big[(1-\eta)J_0(\pi D_g u)+\eta J_0(\pi D_e u)\big].
\end{equation}
Here $\sigma$ is an assumed width and $\eta$ is an assumed relative weight. As an angular plotting scale only, we set $D_g=43.9\,\mu\mathrm{as}$ and $D_e/D_g=1.08719$. The resulting $D_e$ is $47.73\,\mu\mathrm{as}$. These values illustrate the optical ratio, not a fit to EHT visibilities (Fig.~\ref{fig:vis}).

The reason for this reduction is simple. Define $D_{\rm mean}=(1-\eta)D_g+\eta D_e$ and $\Delta D=D_e-D_g$. Expanding a smooth ring response $F(D)$ about $D_{\rm mean}$ gives
\begin{equation}\label{eq:mean-diameter-degeneracy}
\begin{aligned}
 (1-\eta)F(D_g)+\eta F(D_e)
  &=F(D_{\rm mean})\\
  &\quad+\tfrac12\eta(1-\eta)(\Delta D)^2 F''(D_{\rm mean})\\
  &\quad+O((\Delta D)^3).
\end{aligned}
\end{equation}
The linear term cancels. In this toy model, a free mean diameter can therefore absorb much of the leading difference between two nearby optical scales. A real comparison would need polarized radiative transfer, emission parameters, source variability and a likelihood for measured complex visibilities \cite{Broderick2022,Perlick2022,Tlemissov2025,EHTdata,Dahale2025}.

Our calculation gives real-frequency absorption on a static background. Ringdown predictions would also need quasinormal frequencies and an excitation model \cite{BertiCardosoStarinets2009}.

\section{Physical interpretation and relation to earlier work}
The magnetic action determines the geometry and the wave potentials. Changing the charge gives eight black holes within one theory. The separate 21-model optical survey answers a different question: how much can the effective photon orbit change when the geometric capture scale is held fixed across different actions?

The central wave result is the parity difference of the two-channel absorption operator. The larger eigenvalue gives a clear gap near the first absorption step, but its saturation hides the stronger response of the smaller eigenvalue. The trace difference includes both and reaches about $24.3$ percentage points in the extended $\ell=2$ band. The charge-matched RN calculation is a separate comparison with ordinary Einstein--Maxwell gravity. Its parity spectra agree to numerical accuracy on the original band. The RN residuals combine changes in geometry and electromagnetic response, so we do not use them to define the parity gap.

The optical and wave effects probe different aspects of the same matter action. The optical critical scales differ by $3.6$--$7.2\%$. The wave response depends strongly on the extraordinary characteristic coefficient $\kappa$. The formal operator tests help explain this dependence without adding a new source model. Four $\ell=3$ cases support parity-dependent absorption at a higher multipole, but they do not establish a trend for all multipoles.

\subsection{Comparison with the closest literature}
Parity breaking in nonlinear electromagnetic black holes and the possibility of two optical characteristics are known results \cite{Moreno2003,Chaverra2016,DePaula2026}. A recent coupled-wave study of a generalized Hayward magnetic black hole is particularly relevant \cite{Pradhan2026}. Here we use a different magnetic action and keep its parameters linked while varying the black-hole charge. We calculate both optical scales and flux-normalized parity absorption for this family. Table~\ref{tab:literature-r45} summarizes the comparison.

\begin{table*}[!t]
\caption{Position of the present calculation relative to closely related literature. The studies use different sources and compute different observables.}\label{tab:literature-r45}
\centering\small
\begin{tabular}{@{}lll@{}}\toprule
\parbox[t]{.23\textwidth}{\raggedright Study\strut} & \parbox[t]{.30\textwidth}{\raggedright Principal object\strut} & \parbox[t]{.37\textwidth}{\raggedright Distinction from this paper\strut} \\\midrule
\parbox[t]{.23\textwidth}{\raggedright Moreno--Sarbach (2003); Chaverra et al. (2016) \cite{Moreno2003,Chaverra2016}\strut} & \parbox[t]{.30\textwidth}{\raggedright Coupled perturbations and parity splitting in general NLED black-hole families\strut} & \parbox[t]{.37\textwidth}{\raggedright Provides the coupled-wave framework and earlier examples of parity splitting\strut} \\
\parbox[t]{.23\textwidth}{\raggedright De Paula et al. (2026) \cite{DePaula2026}\strut} & \parbox[t]{.30\textwidth}{\raggedright Two characteristic light rings and shadows in a two-invariant NLED model\strut} & \parbox[t]{.37\textwidth}{\raggedright Establishes optical birefringence as prior art; our field action is single invariant with a specified Maxwell-complete Einasto-like core\strut} \\
\parbox[t]{.23\textwidth}{\raggedright Pradhan et al. (2026) \cite{Pradhan2026}\strut} & \parbox[t]{.30\textwidth}{\raggedright Coupled waves of generalized Hayward magnetic NLED black holes\strut} & \parbox[t]{.37\textwidth}{\raggedright Closest operator/scattering comparison; different constitutive law, mass profile and parameter family\strut} \\
\parbox[t]{.23\textwidth}{\raggedright Konoplya--Zhidenko (2026); L\"utf\"uo\u{g}lu et al. (2026) \cite{Konoplya2026,Lutfuoglu2026}\strut} & \parbox[t]{.30\textwidth}{\raggedright Dark-matter-inspired geometries and perturbations\strut} & \parbox[t]{.37\textwidth}{\raggedright Concern different source prescriptions; no identification with the magnetic field action claimed here\strut} \\
\parbox[t]{.23\textwidth}{\raggedright Present paper\strut} & \parbox[t]{.30\textwidth}{\raggedright One magnetic action, two optical scales and flux-normalized parity absorption\strut} & \parbox[t]{.37\textwidth}{\raggedright Calculates the direct parity gap across eight states, with RN as an independent control\strut} \\\bottomrule
\end{tabular}
\end{table*}

\subsection{Physical range of the calculation}
The regular core of this solution is hidden behind the horizon in the cases studied. The exterior has a magnetic power-law tail. We check the matter derivatives, optical characteristic factors and coupled potentials on that exterior. They support the exterior propagation calculation. Global stability would require an additional analysis \cite{Moreno2003,DeFelice2025}.

The eight states are different static solutions of one action. They do not represent the time evolution of a single black hole. The larger absorption eigenvalue describes an optimized coherent incident combination. The fixed gravitational and electromagnetic incident channels, reported separately, are closer to specified scattering experiments. They are not polarization fractions measured in an image. This distinction matters when interpreting the difference between metric parity sectors.

Our $\ell=2$ results cover eight solutions and 31 frequencies. The $\ell=3$ results cover four solutions at higher frequencies. The numerical conclusions apply to these intervals. Extending the multipolar range and connecting absorption to measured signals would require separate calculations.

\section{Conclusion}
We studied two optical characteristics and the absorption of coupled gravitational and electromagnetic waves around a cored magnetic black hole. The mass, charge and core scale follow from one nonlinear electromagnetic action. Eight black holes in this family have optical critical-scale separations of $3.6$--$7.2\%$.

The wave calculation gives different absorption spectra in the two metric-parity sectors. The larger absorption eigenvalue has a gap of about $9.0$ percentage points near its first absorption step. The second eigenvalue carries a stronger gap at higher frequency. Together, the two give a maximum trace difference of about $24.3$ percentage points in the extended $\ell=2$ calculation. The trace is basis independent and has a simple meaning for an equal incoherent mixture of incoming fields.

The wave operator points to a physical origin of the difference: the extraordinary characteristic coefficient $\kappa$ changes the electromagnetic barrier in the even sector. A controlled operator test shows that this coefficient has a large effect on the gap. We also find a parity difference in four $\ell=3$ cases. The charge-matched RN solution provides a separate Einstein--Maxwell control.

The model thus gives a clear example of how one magnetic source can modify both optical propagation and coupled-wave absorption. A direct comparison with astronomical images or ringdown signals would require an emission or waveform model.

\section*{Statements and Declarations}
\noindent\textbf{Funding.} M.F. was supported by the Agencia Nacional de Investigación y Desarrollo (ANID), Chile, through FONDECYT Postdoctorado Grant No. 3260029. F.A. acknowledges the Inter University Centre for Astronomy and Astrophysics (IUCAA), Pune, India, for granting visiting associateship.

\noindent\textbf{Competing interests.} The authors declare no competing interests relevant to this work. 

\noindent\textbf{Data and code availability.} The frequency grids, absorption tables, figure data, and reproducibility notebook are available from the corresponding author upon reasonable request.

\appendix
\section{The even metric-parity operator}\label{app:even}
For full reproducibility we give the canonical even-parity entries used in the code, in the flux normalization of Ref.~\cite{Pradhan2026}. Define
\begin{align}
a&=6m/r-2r^2\mathscr L, &b&=\lambda+4Q^2\mathscr L_{\mathcal F}/r^2,\\
d&=a+\lambda,&c_1&=\lambda+1-f+2r^2\mathscr L,\\
c_2&=c_1+4f\kappa,&w&=c_1+2f\kappa.
\end{align}
Then
\begin{align}
U_+^{gg}&=\frac{L\lambda-2f\lambda+a(a-4m/r)}{r^2d}+\frac{2f\lambda b}{r^2d^2},\\
U_+^{ee}&=\frac{\kappa L}{r^2}+\mathcal D_++\frac{4Q^2\mathscr L_{\mathcal F}c_2}{r^4d}+\frac{8fQ^2\mathscr L_{\mathcal F}b}{r^4d^2},\\
U_+^{ge}&=-\frac{2Q\sqrt{\lambda\mathscr L_{\mathcal F}}}{r^3}\left(\frac wd+\frac{2fb}{d^2}\right).
\end{align}
The full constraint reduction belongs to the cited literature; these equations make the present model-specific substitution reproducible.

\section{Independent source-coefficient checks and reference limits}\label{app:source-validation}
The two-channel operators in this work use the canonical gauge-invariant Einstein--NLED reduction of Refs.~\cite{Moreno2003,Pradhan2026}, specialized to the magnetic source in Eq.~(\ref{eq:profile}). We do not claim an independent derivation of all linearized constraints. For transparency, we provide the following algebraic checks specific to our source.

In the canonical convention $\mathscr L=4\pi\rho$, let $x=(r/R)^q$, $L_F=\mathscr L_{\mathcal F}$ and $A=(\ln L_F)'$, where a prime denotes $d/dr$. Differentiation of our fixed matter action gives
\begin{equation}\label{eq:appendix-source}
\begin{aligned}
 L_F&=\left(\frac{x}{1+x}\right)^{1+4/q},\\
 \kappa&=1-\frac{q+4}{2(1+x)},\\
 A&=\frac{q+4}{r(1+x)},\\
 A'&=-\frac{(q+4)[1+(q+1)x]}{r^2(1+x)^2}.
\end{aligned}
\end{equation}
With $D_- =\sqrt{L_F}\,\partial_r[f\,\partial_r(L_F^{-1/2})]$ and $D_+=L_F^{-1/2}\partial_r[f\,\partial_r(L_F^{1/2})]$, direct differentiation yields the exact identity
\begin{equation}\label{eq:appendix-D}
 D_{\pm}=\frac{f A^2}{4}\pm\frac{f'A+fA'}{2}.
\end{equation}
The source Einstein relation $m'=4\pi r^2\rho$ and the anisotropic pressure $p_t=-\rho-r\rho'/2$ agree with the magnetic NLED expression $2\mathcal F L_F=-2\pi r\rho'$ in this canonical normalization. All source-specific identities above have zero symbolic residual after simplification. The six canonical potential entries used for the two metric-parity sectors were then compared term by term, including the off-diagonal signs in the same master-field basis, with Appendix A of Ref.~\cite{Pradhan2026}; no discrepancy was found.

As independent reference checks, the odd metric-parity gravitational potential reduces to Regge--Wheeler, the even gravitational potential to Zerilli when $Q=0$, and the odd gravitational diagonal to $\ell(\ell+1)/r^2-6M/r^3+4Q^2/r^4$ in the Maxwell--Reissner--Nordstr\"om case. All three reductions have vanishing symbolic residual. We also re-evaluated both matrices from the background without reading precomputed scattering tables for $s=1.08,1.20,1.50$, and checked real symmetry and exterior-spot-point regularity at $r=4M$. These sampled checks neither prove global mode stability nor replace a full first-principles gauge-constraint derivation. The symbolic residuals, sample matrices, and reproducibility notebook are preserved in the project record.

\section{Additional analytic steps for the magnetic completion}\label{app:massdetail}
\subsection{Beta-function normalization and finite-mass test}
Starting from Eq.~(\ref{eq:profile}), the dimensionless substitution $t=(r/R)^q$ gives
\begin{equation}
 \int_0^r\!u^2[1+(u/R)^q]^{-4/q}du
 =\frac{R^3}{q}\int_0^{(r/R)^q}\! t^{3/q-1}(1+t)^{-4/q}dt.
\end{equation}
The second substitution $z=t/(1+t)$ maps the radial integral to the incomplete beta function $B_z(3/q,1/q)$. At infinity its value is $B(3/q,1/q)$. Because both beta arguments are positive for $q>0$, the integral converges and the total ADM mass is finite. This is the analytic basis for choosing the incomplete-beta representation instead of a noisy numerical integral in the orbit and wave solvers. A fresh calculation can verify mass normalization by comparing the beta-function representation with direct quadrature at several radii, including a radius close to the outer horizon.

Differentiating the normalized mass gives $m'=4\pi r^2\rho$ exactly. This identity is especially useful for evaluating $f'$ and the gravitational orbit condition without differentiating a special function numerically. Its second derivative may be written as $m''=8\pi r\rho+4\pi r^2\rho'$. With $\rho'=-(4x/r(1+x))\rho$, both derivatives are elementary in $x$ and are regular for $r>0$. These formulas also expose the distinct limits at small and large radius: $m\sim r^3$ near the center and $M-m\sim r^{-1}$ in the Maxwell far field.

\subsection{A direct check of the kinetic coefficient}
From $\mathcal F=Q^2/(2r^4)$ and $\rho=\rho_0(1+x)^{-4/q}$ one obtains
\begin{equation}
\frac{d\rho}{dr}=-\frac{4x}{r(1+x)}\rho,
\qquad\frac{d\mathcal F}{dr}=-\frac{4\mathcal F}{r}.
\end{equation}
Their ratio is $\mathcal L_{\mathcal F}=\rho x/[\mathcal F(1+x)]$. Since $\rho=\rho_0(1+x)^{-4/q}$ and $\mathcal F=\mathcal F_0x^{-4/q}$, this becomes
\begin{equation}
 \mathcal L_{\mathcal F}=\frac{\rho_0}{\mathcal F_0}\left(\frac{x}{1+x}\right)^{1+4/q}.
\end{equation}
The Maxwell condition $\mathcal L_{\mathcal F}\to1/(4\pi)$ therefore fixes $\mathcal F_0=4\pi\rho_0$, consistent with the asymptotic gravitational charge. Differentiating once more in $\mathcal F$ and forming $1+2\mathcal F\mathcal L_{\mathcal F\mathcal F}/\mathcal L_{\mathcal F}$ yields Eq.~(\ref{eq:kappa}). This derivation does not refer to the numerical wave matrices and is therefore a separate check of the local constitutive characteristic factor.

\subsection{Regularity is not the same as stability}
Equation~(\ref{eq:deSitter-core}) establishes finite background curvature near $r=0$ for the parameter range considered. The action derivative, however, behaves differently there: $\mathcal L_{\mathcal F}$ decreases toward the center while $\kappa$ can become negative. An exterior event horizon hides this region from the scattering domain, but hiding it is not the same as proving well-posedness of the complete nonlinear initial-value problem. This is one reason why existing no-go and instability analyses for NLED regular centers are cited directly \cite{Bokulic2022,Bokulic2026,DeFelice2025}. It would be incorrect to extrapolate the positive exterior numerical matrix eigenvalues into a proof of global regular stability.

\section{Optical and wave diagnostic definitions}\label{app:definitions}
\subsection{The relation between bending slope and orbit instability}
For a general characteristic potential $U_j=fK_j/r^2$, define $\Omega_j=\sqrt{U_j(r_j)}$ and the local coordinate-time Lyapunov exponent
\begin{equation}
\Lambda_j=f(r_j)\sqrt{-\frac{U_j''(r_j)}{2U_j(r_j)}}.
\end{equation}
Inserting these definitions into Eq.~(\ref{eq:strong}) gives the useful identity
\begin{equation}\label{eq:bozza-lyapunov}
 \bar a_j=\frac{\Omega_j}{\Lambda_j}.
\end{equation}
Indeed, $U_j(r_j)=f(r_j)K_j(r_j)/r_j^2$, so the factors cancel algebraically. This identity is a cross-check between the strong-deflection coefficient obtained from the ray integral and the local second derivative of the radial characteristic potential. It does not turn the extraordinary electromagnetic orbit into a gravitational quasinormal frequency; for that one must check the principal symbol of the relevant coupled wave branch. The relation is used only within the stated static characteristic Hamiltonian model.

\subsection{A precise interpretation of the finite-band mean}
Let $x_\omega=M\omega$ and $I=[0.30,0.60]$. Our uniform average is
\begin{equation}
 \overline{\delta\Gamma}_{\mathcal P}(s)=\frac{1}{|I|}\int_I\delta\Gamma_{\mathcal P}(s,x_\omega)dx_\omega.
\end{equation}
A measured frequency average would depend on a source and detector weight $W(x_\omega)$, with the form $\int W(x_\omega)\delta\Gamma(x_\omega)dx_\omega/\int Wdx_\omega$. We use a uniform weight to compare the eight model states. Zero crossings are checked by direct wave integrations and changes to the numerical boundaries.

\subsection{Optical, scattering and observational quantities measure different things}
The results describe four distinct quantities. Critical impact parameters follow from maxima of the characteristic potentials. Bending angles come from ray integrals. Absorption probabilities come from the coupled-channel flux matrix. The illustrative visibility comes from a simple ring model. Keeping these quantities separate makes the physical meaning of each result clear.

\section{Numerical dataset and independent reproduction protocol}\label{app:numerical-reproduction}
\subsection{Fixed-action state and frequency grid}
The data tables give $s$, $Q/M$, the two optical impact parameters and the absorption probabilities for each frequency and parity. We also retain the ten-frequency RN comparison as a separate control. The eight-state grid is $s\in\{1.08,1.10,1.15,1.20,1.25,1.30,1.40,1.50\}$. The wave records store the four independent labels (model, parity, state, frequency) before taking differences. This matters because an incorrectly matched state, frequency, or metric parity could generate a smooth-looking but spurious residual. We cross-check by joining rows on these labels, not by subtracting two arrays on the assumption that their row order matches.

At a fixed $s$, the asymptotic RN charge is read from the same metric coefficient that determines the NLED magnetic charge. We use the same dimensionless $M\omega$ for both models, and both parity computations are compared at the same frequency. The tables and maps are calculated from the stored scattering solutions. The archived records include the full input labels and the calculated probabilities. They allow the figures and summary tables to be checked without reading a rendered plot.

\subsection{Minimal reproducibility sequence}
To reproduce the wave calculation, we use the following steps. Fix $q$ and the reference action constants; choose $s$ and obtain $R(s),M(s),Q(s)$ from Eq.~(\ref{eq:fixed-action}). Compute $m(r)$ with the incomplete beta function and find the largest positive root of $f$. Check that $r_h>r_\kappa$ and build the two optical characteristic potentials. Locate their exterior unstable maxima by derivative root search; calculate $r_j,b_j$ and optionally the bending integral. Then evaluate the canonical source coefficients $\mathscr L_{\mathcal F},\kappa,A,A'$ from their analytical expressions rather than finite-differencing them. Form the odd and even matrix potentials at $\ell=2$ and integrate two independent ingoing horizon solutions for each real frequency.

At the outer matching surface, extract the two incoming and outgoing amplitude matrices in the same flux normalization and form $\mathbf R$ and $\mathbf T$. Evaluate the eigenvalues of $\mathbf T^\dagger\mathbf T$, check their range and the matrix flux identity. Repeat the operation on the RN black-hole comparator with the same $M$ and $Q$. Subtract the higher-eigenvalue channel eigenvalues at identical states and frequencies. The parity crossing requires a separate bracketed real-frequency root; its value cannot be read reliably from the ten-point grid alone. Numerical confidence requires independent variation of the outer boundary, the near-horizon start and the frequency quadrature, because those control different numerical errors.

An RN black-hole reference requires $Q/M<1$. With the exact fixed-action scaling $Q/M=(Q_\star/M_\star)s^{-1/2}$, the RN extremality threshold is $s>(Q_\star/M_\star)^2\simeq1.06684$. The eight-state grid begins at $s=1.08$ and thus lies on the permitted side; the $s=1$ NLED solution remains meaningful but has no subextremal RN black-hole counterpart. This threshold can be checked independently of any wave integration and should be tested before running the 320-case scattering scan.

\subsection{Numerical records and cross-checks}
The wave records contain the parameters, frequencies and flux-normalized probabilities. The published tables can be reconstructed from these records. Separate wave integrations were used to test the frequency grid, the numerical boundaries and selected solutions with a second solver. The two checks address different sources of numerical error.

\section{Full radial and scattering summaries}\label{app:full-data}
\subsection{The 27-state exterior optical survey}
Table~\ref{tab:fulloptical} gives the full fixed-action exterior optical sweep used to establish the allowed region and the trend of both characteristic impact parameters. These states share the same $q=3/2$ constitutive action and differ only through the monopole charge label $s$ and the consequent mass and size scaling. The table is a deterministic model survey, not a posterior or a population sample. It makes the precise relationship between the exploratory 27-state sequence and the RN-comparable eight-state wave subset visible to the reader. In particular, a positive exterior $\kappa$ does not by itself guarantee that an RN comparator has an event horizon.

\begin{center}\begin{minipage}{\columnwidth}\centering\small
\captionof{table}{Full 27-state exterior optical survey of the fixed-action family ($q=3/2$).}\label{tab:fulloptical}
\begin{tabular}{@{}rrrrr@{}}\toprule
$s$ & $r_h/M$ & $b_{\rm g}/M$ & $b_{\rm e}/M$ & Split (\%) \\\midrule
0.850 & 0.959 & 3.666 & 4.180 & 14.01 \\
0.875 & 1.034 & 3.737 & 4.213 & 12.73 \\
0.900 & 1.092 & 3.801 & 4.244 & 11.67 \\
0.925 & 1.139 & 3.858 & 4.273 & 10.76 \\
0.950 & 1.180 & 3.909 & 4.300 & 9.99 \\
0.975 & 1.216 & 3.957 & 4.325 & 9.31 \\
1.000 & 1.248 & 4.000 & 4.349 & 8.72 \\
1.025 & 1.277 & 4.040 & 4.371 & 8.19 \\
1.050 & 1.304 & 4.078 & 4.392 & 7.72 \\
1.075 & 1.328 & 4.112 & 4.412 & 7.30 \\
1.100 & 1.351 & 4.145 & 4.431 & 6.91 \\
1.125 & 1.371 & 4.175 & 4.449 & 6.56 \\
1.150 & 1.391 & 4.204 & 4.467 & 6.25 \\
1.175 & 1.409 & 4.231 & 4.483 & 5.95 \\
1.200 & 1.426 & 4.256 & 4.499 & 5.69 \\
1.225 & 1.442 & 4.281 & 4.513 & 5.44 \\
1.250 & 1.457 & 4.303 & 4.528 & 5.21 \\
1.275 & 1.471 & 4.325 & 4.541 & 5.00 \\
1.300 & 1.484 & 4.346 & 4.554 & 4.80 \\
1.325 & 1.497 & 4.365 & 4.567 & 4.61 \\
1.350 & 1.509 & 4.384 & 4.579 & 4.44 \\
1.375 & 1.520 & 4.402 & 4.590 & 4.28 \\
1.400 & 1.531 & 4.419 & 4.601 & 4.13 \\
1.425 & 1.541 & 4.435 & 4.612 & 3.98 \\
1.450 & 1.551 & 4.451 & 4.622 & 3.85 \\
1.475 & 1.561 & 4.466 & 4.632 & 3.72 \\
1.500 & 1.570 & 4.480 & 4.642 & 3.60 \\
\bottomrule\end{tabular}
\end{minipage}\end{center}
\noindent\textit{Reading the table.} The geometric and extraordinary critical parameters are computed from two different radial potentials on each background, not through an assumed proportionality. The optical separation shrinks over the surveyed branch, and both critical parameters are normalized to the ADM mass of their own state. The background central density and electromagnetic normalization were held fixed by the action. The wave tables below use only the eight states with a regular RN black-hole control.

\subsection{All ten sampled frequencies for the wave subset}
For transparency, the ten-frequency results are reported here as signed RN-subtracted absorption eigenvalue differences, multiplied by one hundred so their numerical units are percentage points. The two parity columns are calculated from separate coupled operators. These tables record the eight-state, two-parity, charge-matched calculations. Direct zero-crossing frequencies need not coincide with one of these ten frequencies because they were computed by a separate bracketed solver. The results for $s=1.08$, $1.10$, $1.15$, and $1.20$ appear in Tables~\ref{tab:wave-residual-108}, \ref{tab:wave-residual-110}, \ref{tab:wave-residual-115}, and \ref{tab:wave-residual-120}, respectively. Those for $s=1.25$, $1.30$, $1.40$, and $1.50$ appear in Tables~\ref{tab:wave-residual-125}, \ref{tab:wave-residual-130}, \ref{tab:wave-residual-140}, and \ref{tab:wave-residual-150}, respectively.

\begin{center}\begin{minipage}{\columnwidth}\centering\small
\captionof{table}{RN-subtracted wave absorption residuals for $s=1.08$ and $Q/M=0.9939$ (percentage points).}\label{tab:wave-residual-108}
\begin{tabular}{@{}rrr@{}}\toprule
$M\omega$ & Odd residual (pp) & Even residual (pp) \\\midrule
0.30 & +0.239 & +0.561 \\
0.34 & +0.771 & +2.191 \\
0.38 & +1.719 & +6.544 \\
0.40 & +1.920 & +9.197 \\
0.42 & +1.472 & +10.360 \\
0.45 & +0.009 & +7.608 \\
0.48 & -0.696 & +3.442 \\
0.52 & -0.473 & +0.814 \\
0.56 & -0.186 & +0.164 \\
0.60 & -0.062 & +0.031 \\
\bottomrule\end{tabular}\end{minipage}\end{center}
\begin{center}\begin{minipage}{\columnwidth}\centering\small
\captionof{table}{RN-subtracted wave absorption residuals for $s=1.10$ and $Q/M=0.9848$ (percentage points).}\label{tab:wave-residual-110}
\begin{tabular}{@{}rrr@{}}\toprule
$M\omega$ & Odd residual (pp) & Even residual (pp) \\\midrule
0.30 & +0.223 & +0.545 \\
0.34 & +0.710 & +2.121 \\
0.38 & +1.539 & +6.255 \\
0.40 & +1.670 & +8.682 \\
0.42 & +1.213 & +9.627 \\
0.45 & -0.084 & +6.931 \\
0.48 & -0.646 & +3.119 \\
0.52 & -0.422 & +0.745 \\
0.56 & -0.166 & +0.153 \\
0.60 & -0.056 & +0.029 \\
\bottomrule\end{tabular}\end{minipage}\end{center}
\begin{center}\begin{minipage}{\columnwidth}\centering\small
\captionof{table}{RN-subtracted wave absorption residuals for $s=1.15$ and $Q/M=0.9632$ (percentage points).}\label{tab:wave-residual-115}
\begin{tabular}{@{}rrr@{}}\toprule
$M\omega$ & Odd residual (pp) & Even residual (pp) \\\midrule
0.30 & +0.188 & +0.507 \\
0.34 & +0.581 & +1.959 \\
0.38 & +1.166 & +5.601 \\
0.40 & +1.168 & +7.548 \\
0.42 & +0.722 & +8.076 \\
0.45 & -0.232 & +5.568 \\
0.48 & -0.543 & +2.475 \\
0.52 & -0.323 & +0.603 \\
0.56 & -0.126 & +0.128 \\
0.60 & -0.043 & +0.026 \\
\bottomrule\end{tabular}\end{minipage}\end{center}
\begin{center}\begin{minipage}{\columnwidth}\centering\small
\captionof{table}{RN-subtracted wave absorption residuals for $s=1.20$ and $Q/M=0.9429$ (percentage points).}\label{tab:wave-residual-120}
\begin{tabular}{@{}rrr@{}}\toprule
$M\omega$ & Odd residual (pp) & Even residual (pp) \\\midrule
0.30 & +0.160 & +0.474 \\
0.34 & +0.477 & +1.813 \\
0.38 & +0.881 & +5.036 \\
0.40 & +0.805 & +6.606 \\
0.42 & +0.395 & +6.848 \\
0.45 & -0.306 & +4.550 \\
0.48 & -0.462 & +2.004 \\
0.52 & -0.255 & +0.496 \\
0.56 & -0.098 & +0.109 \\
0.60 & -0.034 & +0.023 \\
\bottomrule\end{tabular}\end{minipage}\end{center}
\begin{center}\begin{minipage}{\columnwidth}\centering\small
\captionof{table}{RN-subtracted wave absorption residuals for $s=1.25$ and $Q/M=0.9238$ (percentage points).}\label{tab:wave-residual-125}
\begin{tabular}{@{}rrr@{}}\toprule
$M\omega$ & Odd residual (pp) & Even residual (pp) \\\midrule
0.30 & +0.136 & +0.443 \\
0.34 & +0.393 & +1.682 \\
0.38 & +0.663 & +4.545 \\
0.40 & +0.542 & +5.817 \\
0.42 & +0.177 & +5.865 \\
0.45 & -0.338 & +3.776 \\
0.48 & -0.397 & +1.648 \\
0.52 & -0.206 & +0.414 \\
0.56 & -0.079 & +0.093 \\
0.60 & -0.027 & +0.020 \\
\bottomrule\end{tabular}\end{minipage}\end{center}
\begin{center}\begin{minipage}{\columnwidth}\centering\small
\captionof{table}{RN-subtracted wave absorption residuals for $s=1.30$ and $Q/M=0.9059$ (percentage points).}\label{tab:wave-residual-130}
\begin{tabular}{@{}rrr@{}}\toprule
$M\omega$ & Odd residual (pp) & Even residual (pp) \\\midrule
0.30 & +0.117 & +0.415 \\
0.34 & +0.324 & +1.563 \\
0.38 & +0.494 & +4.116 \\
0.40 & +0.349 & +5.152 \\
0.42 & +0.031 & +5.069 \\
0.45 & -0.346 & +3.176 \\
0.48 & -0.345 & +1.376 \\
0.52 & -0.170 & +0.349 \\
0.56 & -0.064 & +0.080 \\
0.60 & -0.022 & +0.018 \\
\bottomrule\end{tabular}\end{minipage}\end{center}
\begin{center}\begin{minipage}{\columnwidth}\centering\small
\captionof{table}{RN-subtracted wave absorption residuals for $s=1.40$ and $Q/M=0.8729$ (percentage points).}\label{tab:wave-residual-140}
\begin{tabular}{@{}rrr@{}}\toprule
$M\omega$ & Odd residual (pp) & Even residual (pp) \\\midrule
0.30 & +0.086 & +0.366 \\
0.34 & +0.221 & +1.357 \\
0.38 & +0.261 & +3.412 \\
0.40 & +0.104 & +4.107 \\
0.42 & -0.131 & +3.877 \\
0.45 & -0.326 & +2.325 \\
0.48 & -0.266 & +0.995 \\
0.52 & -0.121 & +0.256 \\
0.56 & -0.045 & +0.061 \\
0.60 & -0.015 & +0.014 \\
\bottomrule\end{tabular}\end{minipage}\end{center}
\begin{center}\begin{minipage}{\columnwidth}\centering\small
\captionof{table}{RN-subtracted wave absorption residuals for $s=1.50$ and $Q/M=0.8433$ (percentage points).}\label{tab:wave-residual-150}
\begin{tabular}{@{}rrr@{}}\toprule
$M\omega$ & Odd residual (pp) & Even residual (pp) \\\midrule
0.30 & +0.064 & +0.324 \\
0.34 & +0.150 & +1.186 \\
0.38 & +0.118 & +2.864 \\
0.40 & -0.029 & +3.336 \\
0.42 & -0.199 & +3.045 \\
0.45 & -0.291 & +1.765 \\
0.48 & -0.211 & +0.748 \\
0.52 & -0.091 & +0.195 \\
0.56 & -0.033 & +0.047 \\
0.60 & -0.011 & +0.011 \\
\bottomrule\end{tabular}\end{minipage}\end{center}
\noindent A reader can reproduce the band averages by applying the stated trapezoidal rule to each state--parity column, multiplying back by the appropriate $M\omega$ interval and dividing by its width. Those ten-point values and the direct zero-crossing checks should be compared with the denser 19-point endpoint calculation before interpreting the cancellation in the odd sector. The full precision numerical columns, flux residuals and two-channel eigenvalues are preserved as CSV files in the companion research archive and in the self-contained regeneration notebook.

\section{Additional optical and multipolar checks}\label{app:secondary-grid}
\subsection{The matched-geometric control: all 21 constitutive models}
The main manuscript reports three illustrative members of the 21-background comparison in which the geometric critical impact parameter is held at $4M$. Table~\ref{tab:app-matched-21} permits a direct check of how this constraint affects the extraordinary critical parameter, the photon-orbit instability and both logarithmic strong-deflection coefficients. This study deliberately compares \emph{different} magnetic matter actions because $q$ and $R$ change together to maintain the geometric capture scale. It does not represent an eight-state or 27-state fixed-action sequence. In particular, no claim of a continuously adjustable magnetic source within one matter theory follows from this matching experiment.

\begin{center}\begin{minipage}{\columnwidth}\centering\footnotesize
\captionof{table}{Matched-geometric control: optical properties of all 21 magnetic models.}\label{tab:app-matched-21}
\begin{tabular}{@{}rrrrr@{}}\toprule
$q$ & $R/M$ & $b_e/M$ & $M\Lambda_g$ & $\bar a_e$ \\\midrule
1.00 & 0.2005 & 4.5142 & 0.20117 & 1.1327 \\
1.05 & 0.2143 & 4.4945 & 0.20006 & 1.1417 \\
1.10 & 0.2278 & 4.4756 & 0.19900 & 1.1502 \\
1.15 & 0.2408 & 4.4575 & 0.19799 & 1.1582 \\
1.20 & 0.2533 & 4.4401 & 0.19702 & 1.1658 \\
1.25 & 0.2655 & 4.4233 & 0.19609 & 1.1731 \\
1.30 & 0.2772 & 4.4072 & 0.19520 & 1.1800 \\
1.35 & 0.2885 & 4.3917 & 0.19434 & 1.1865 \\
1.40 & 0.2994 & 4.3769 & 0.19353 & 1.1927 \\
1.45 & 0.3099 & 4.3625 & 0.19275 & 1.1987 \\
1.50 & 0.3201 & 4.3488 & 0.19200 & 1.2043 \\
1.55 & 0.3298 & 4.3355 & 0.19128 & 1.2097 \\
1.60 & 0.3392 & 4.3227 & 0.19060 & 1.2149 \\
1.65 & 0.3483 & 4.3104 & 0.18994 & 1.2199 \\
1.70 & 0.3570 & 4.2986 & 0.18931 & 1.2246 \\
1.75 & 0.3654 & 4.2872 & 0.18871 & 1.2292 \\
1.80 & 0.3735 & 4.2762 & 0.18814 & 1.2335 \\
1.85 & 0.3813 & 4.2657 & 0.18759 & 1.2377 \\
1.90 & 0.3888 & 4.2555 & 0.18707 & 1.2418 \\
1.95 & 0.3961 & 4.2457 & 0.18657 & 1.2457 \\
2.00 & 0.4031 & 4.2363 & 0.18609 & 1.2495 \\
\bottomrule\end{tabular}\end{minipage}\end{center}
The trend of $M\Lambda_g$ is a direct demonstration that a fixed \emph{value} of a potential at its maximum does not fix its local \emph{curvature}. The relation $\bar a_g=\Omega_g/\Lambda_g$ from Appendix~\ref{app:definitions} supplies a second analytic check on the near-critical bending coefficient computed from the strong-deflection integral. The table contains directly evaluated optical quantities. No spectral fit is used.

\subsection{A bounded check beyond the quadrupole}\label{app:l3}
The central wave analysis uses $\ell=2$, where the coupled gravitational--electromagnetic problem is physically relevant and the full fixed-action RN subtraction has been tested. The two-frequency $\ell=3$ spot check is reported in Table~\ref{tab:app-l3-spot} for comparison with the broader four-state frequency test in Sec.~\ref{sec:freq-refinement}. It checks that the numerical operator construction and the channel normalization are not hard-coded to the quadrupole. The expanded test examines direct parity absorption, not whether the quadrupolar RN-subtracted zero crossing persists at every multipole.

\begin{center}\begin{minipage}{\columnwidth}\centering\footnotesize
\captionof{table}{Two-frequency $\ell=3$ absorption and flux checks.}\label{tab:app-l3-spot}
\begin{tabular}{@{}rrrrr@{}}\toprule
$s$ & $M\omega$ & Metric parity & $\Gamma_{\rm high}$ & Flux error \\\midrule
1.0 & 0.45 & odd & 0.00006 & 1.2e-11 \\
1.0 & 0.45 & even & 0.00016 & 3.4e-11 \\
1.0 & 0.65 & odd & 0.09508 & 3.1e-08 \\
1.0 & 0.65 & even & 0.21022 & 5.7e-08 \\
1.2 & 0.45 & odd & 0.00024 & 4.9e-11 \\
1.2 & 0.45 & even & 0.00041 & 6.5e-11 \\
1.2 & 0.65 & odd & 0.25001 & 7.9e-08 \\
1.2 & 0.65 & even & 0.35782 & 9.8e-08 \\
1.5 & 0.45 & odd & 0.00063 & 1.3e-10 \\
1.5 & 0.45 & even & 0.00084 & 1.5e-10 \\
1.5 & 0.65 & odd & 0.43616 & 1.4e-07 \\
1.5 & 0.65 & even & 0.50554 & 1.4e-07 \\
\bottomrule\end{tabular}\end{minipage}\end{center}
The $\ell=3$ probabilities show the expected sensitivity of a higher angular momentum barrier to frequency: in the displayed states, the higher-eigenvalue channel absorption is strongly suppressed at $M\omega=0.45$ and becomes much larger by $M\omega=0.65$. The two metric parity sectors need not be identical for this nonlinear source. Because the small sample contains no complete RN subtraction across a continuum of frequencies, we draw no claim about an $\ell=3$ signed residual root. The conserved-flux checks remain small in each case. This bounded multipole test is a validation supplement, not an additional primary observable.

\section{Reconstruction of the joint diagnostic}\label{app:r46derived}
The eight-state joint diagnostic uses the fixed-action sequence. The optical split is computed from the two critical impact parameters. The wave peak and its location are drawn from the locally optimized $\ell=2$ scattering result, and the band mean is computed from the 31-point direct parity-difference grid. Thus each plotted point has a defined numerical origin.

The parity band mean is
\begin{equation}
\overline{\Delta\Gamma}_{\rm par}(s)=\frac{1}{0.30}\int_{0.30}^{0.60}\Delta\Gamma_{\rm par}(s,\omega)\,d(M\omega),
\end{equation}
with the integral evaluated by the trapezoidal rule. For $s=1.08$ it is about $3.34$ percentage points; for $s=1.50$ it is about $1.27$ percentage points. The ten-point survey and the selected 19-point re-integrations provide independent checks of the band integration. The main table uses the same 31-point rule for all states.

The local peak values and their frequencies are listed in Table~\ref{tab:R53-trace-peaks}. They follow from a deterministic parameter survey.

\section{An exact optical response identity for a fixed-action sequence}\label{app:envelope}
The optical table and the joint-diagnostic plane have an analytic
consistency relation that does not require a fit or another ray
integration. It also clarifies which part of the variation comes from
moving the critical radius and which part comes from changing the
underlying optical potential. Because the ADM mass varies across the
fixed-action family, it is essential to use the dimensionless radius
$x=r/M(s)$ before taking a derivative with respect to the family
parameter $s$. In this normalization write
\begin{equation}
 h_\alpha(x,s)=\frac{f(x,s)\alpha(x,s)}{x^2},
 \quad \alpha_{\gmode}=1,\quad \alpha_{\ee}=\kappa.
\end{equation}
For each polarization, the unstable optical circular orbit $x_\alpha(s)$
is a nondegenerate local maximum of $h_\alpha$ outside the horizon,
with $\partial_x h_\alpha(x_\alpha,s)=0$. The critical impact parameter
in ADM-mass units is $B_\alpha(s)=b_\alpha/M=h_\alpha(x_\alpha,s)^{-1/2}$.
A total derivative then gives
\begin{align}
 \frac{d\ln B_\alpha}{ds}
 &=-\frac12\left[
 \partial_s\ln h_\alpha
 +\partial_x\ln h_\alpha\,\frac{dx_\alpha}{ds}
 \right]_{x=x_\alpha(s)}\\
 &=-\frac12\left.\partial_s\ln h_\alpha\right|_{x=x_\alpha(s)}.
 \label{eq:envelope}
\end{align}
The derivative of the critical radius cancels \emph{exactly} because
that radius is stationary for the same Hamiltonian whose critical
impact parameter is being evaluated. Thus, the logarithmic response of
the ratio of the two critical scales is
\begin{equation}
 \frac{d}{ds}\ln\!\left(\frac{B_{\ee}}{B_{\gmode}}\right)
 =-\frac12\left[
 \left.\partial_s\ln h_{\ee}\right|_{x_{\ee}}
 -\left.\partial_s\ln h_{\gmode}\right|_{x_{\gmode}}
 \right].\label{eq:envelope-ratio}
\end{equation}
This envelope identity is independent of any approximation that equates
the two critical radii. It is also not a claim that the radii are
unchanged as $s$ varies. Both $x_{\gmode}$ and $x_{\ee}$ generally move;
Eq.~(\ref{eq:envelope}) says only that their first-order displacements
do not enter the derivative of their \emph{stationary values}. In
particular, replacing $x_{\ee}$ by $x_{\gmode}$ in the first term of
Eq.~(\ref{eq:envelope-ratio}) would discard part of the constitutive
effect and is not justified.

The identity offers a clean extension path: with differentiable
interpolants of the exact normalized background $f(x,s)$ and
constitutive factor $\kappa(x,s)$, its right-hand side can be checked
against centered finite differences of the 27-state optical sequence.
We do not promote that proposed additional check to a completed
continuum calculation in this manuscript. The available discrete
states already fix the direction and magnitude of the change in
$B_{\ee}/B_{\gmode}$, while the eight-state joint panel shows that the
wave-contrast-to-optical-gap ratio is not constant. Equation
(\ref{eq:envelope-ratio}) explains the geometrical half of that
comparison without asserting a universal functional relation to a
finite-frequency scattering operator. The equation assumes smooth
nondegenerate maxima and need not hold unchanged through a photon
orbit merger, a zero of $\kappa$, or a change of the physical
exterior branch. The numerical parameter range used here avoids such
an extrapolation, consistent with the strong-deflection and optical
conditions stated in the main text
\cite{Bozza2002,Perlick2022,RussoTownsend2026}.

\bibliographystyle{unsrt}
\bibliography{references}

@article{Novello2000,
  author = {M. Novello and V. A. De Lorenci and J. M. Salim and R. Klippert},
  title = {{Geometrical aspects of light propagation in nonlinear electrodynamics}},
  journal = {Phys. Rev. D},
  volume = {61},
  pages = {045001},
  year = {2000},
  doi = {10.1103/PhysRevD.61.045001},
  note = {\href{https://doi.org/10.1103/PhysRevD.61.045001}{doi:10.1103/PhysRevD.61.045001}}
}

@article{Moreno2003,
  author = {C. Moreno and O. Sarbach},
  title = {{Stability properties of black holes in self-gravitating nonlinear electrodynamics}},
  journal = {Phys. Rev. D},
  volume = {67},
  pages = {024028},
  year = {2003},
  doi = {10.1103/PhysRevD.67.024028},
  note = {\href{https://doi.org/10.1103/PhysRevD.67.024028}{doi:10.1103/PhysRevD.67.024028}}
}

@article{Walia2024,
  author = {R. K. Walia},
  title = {{Exploring nonlinear electrodynamics theories: Shadows of regular black holes and horizonless ultracompact objects}},
  journal = {Phys. Rev. D},
  volume = {110},
  pages = {064058},
  year = {2024},
  doi = {10.1103/PhysRevD.110.064058},
  note = {\href{https://doi.org/10.1103/PhysRevD.110.064058}{doi:10.1103/PhysRevD.110.064058}}
}

@article{AyonBeato1998,
  author = {E. Ay\'on-Beato and A. Garc\'ia},
  title = {{Regular black hole in general relativity coupled to nonlinear electrodynamics}},
  journal = {Phys. Rev. Lett.},
  volume = {80},
  pages = {5056},
  year = {1998},
  doi = {10.1103/PhysRevLett.80.5056},
  note = {\href{https://doi.org/10.1103/PhysRevLett.80.5056}{doi:10.1103/PhysRevLett.80.5056}}
}

@article{Bronnikov2001,
  author = {K. A. Bronnikov},
  title = {{Regular magnetic black holes and monopoles from nonlinear electrodynamics}},
  journal = {Phys. Rev. D},
  volume = {63},
  pages = {044005},
  year = {2001},
  doi = {10.1103/PhysRevD.63.044005},
  note = {\href{https://doi.org/10.1103/PhysRevD.63.044005}{doi:10.1103/PhysRevD.63.044005}}
}

@article{ReggeWheeler1957,
  author = {T. Regge and J. A. Wheeler},
  title = {{Stability of a Schwarzschild singularity}},
  journal = {Phys. Rev.},
  volume = {108},
  pages = {1063},
  year = {1957},
  doi = {10.1103/PhysRev.108.1063},
  note = {\href{https://doi.org/10.1103/PhysRev.108.1063}{doi:10.1103/PhysRev.108.1063}}
}

@article{Zerilli1970,
  author = {F. J. Zerilli},
  title = {{Effective potential for even-parity Regge--Wheeler gravitational perturbation equations}},
  journal = {Phys. Rev. Lett.},
  volume = {24},
  pages = {737},
  year = {1970},
  doi = {10.1103/PhysRevLett.24.737},
  note = {\href{https://doi.org/10.1103/PhysRevLett.24.737}{doi:10.1103/PhysRevLett.24.737}}
}

@article{Chaverra2016,
  author = {E. Chaverra and J. C. Degollado and C. Moreno and O. Sarbach},
  title = {{Black holes in nonlinear electrodynamics: Quasinormal spectra and parity splitting}},
  journal = {Phys. Rev. D},
  volume = {93},
  pages = {123013},
  year = {2016},
  doi = {10.1103/PhysRevD.93.123013},
  note = {\href{https://doi.org/10.1103/PhysRevD.93.123013}{doi:10.1103/PhysRevD.93.123013}}
}

@article{Nomura2022,
  author = {K. Nomura and D. Yoshida},
  title = {{Quasinormal modes of charged black holes with corrections from nonlinear electrodynamics}},
  journal = {Phys. Rev. D},
  volume = {105},
  pages = {044006},
  year = {2022},
  doi = {10.1103/PhysRevD.105.044006},
  note = {\href{https://doi.org/10.1103/PhysRevD.105.044006}{doi:10.1103/PhysRevD.105.044006}}
}

@article{DePaula2026,
  author = {M. A. A. de Paula and H. C. D. Lima and P. V. P. Cunha and C. A. R. Herdeiro and L. C. B. Crispino},
  title = {{Two shadows of a single black hole: Vacuum birefringence phenomena within Einstein-nonlinear-electrodynamics}},
  journal = {Phys. Rev. D},
  volume = {114},
  pages = {024043},
  year = {2026},
  doi = {10.1103/8lkf-mgjw},
  note = {\href{https://doi.org/10.1103/8lkf-mgjw}{doi:10.1103/8lkf-mgjw}}
}

@article{Pradhan2026,
  author = {A. Pradhan and K. Ghaderi and M. Zeyauddin and A. Gulhane},
  title = {{Coupled gravitoelectromagnetic response of a magnetically supported generalized Hayward black hole in nonlinear electrodynamics}},
  journal = {Eur. Phys. J. C},
  volume = {86},
  pages = {1140},
  year = {2026},
  doi = {10.1140/epjc/s10052-026-16428-y},
  note = {\href{https://doi.org/10.1140/epjc/s10052-026-16428-y}{doi:10.1140/epjc/s10052-026-16428-y}}
}

@article{Daghigh2022,
  author = {R. G. Daghigh and M. D. Green},
  title = {{Gravitational and electromagnetic radiation from an electrically charged black hole in general nonlinear electrodynamics}},
  journal = {Phys. Rev. D},
  volume = {105},
  pages = {024055},
  year = {2022},
  doi = {10.1103/PhysRevD.105.024055},
  note = {\href{https://doi.org/10.1103/PhysRevD.105.024055}{doi:10.1103/PhysRevD.105.024055}}
}

@article{NLEDQNM2026,
  author = {M. Fathi and A. Guzm\'an and J. R. Villanueva},
  title = {{Quasinormal modes of a static black hole in nonlinear electrodynamics}},
  journal = {Eur. Phys. J. C},
  volume = {86},
  pages = {587},
  year = {2026},
  doi = {10.1140/epjc/s10052-026-15713-0},
  note = {\href{https://doi.org/10.1140/epjc/s10052-026-15713-0}{doi:10.1140/epjc/s10052-026-15713-0}}
}

@article{Konoplya2026,
  author = {R. A. Konoplya and A. Zhidenko},
  title = {{Dark matter halo as a source of regular black-hole geometries}},
  journal = {Phys. Rev. D},
  volume = {113},
  pages = {043011},
  year = {2026},
  doi = {10.1103/7ptp-9j1t},
  note = {\href{https://doi.org/10.1103/7ptp-9j1t}{doi:10.1103/7ptp-9j1t}}
}

@article{Lutfuoglu2026,
  author = {B. C. L\"utf\"uo\u{g}lu and J. Rayimbaev and S. Murodov and M. Abdullaev and M. Akhmedov},
  title = {{Ringing regularity: Gravitational perturbations and quasinormal modes of Einasto-supported black holes}},
  journal = {Ann. Phys.},
  volume = {494},
  pages = {170664},
  year = {2026},
  doi = {10.1016/j.aop.2026.170664},
  note = {\href{https://doi.org/10.1016/j.aop.2026.170664}{doi:10.1016/j.aop.2026.170664}}
}

@article{Bolokhov2026,
  author = {S. V. Bolokhov},
  title = {{Revisiting black holes in dark-matter halos: on consistent solutions to the Einstein equations}},
  journal = {Eur. Phys. J. C},
  volume = {86},
  pages = {576},
  year = {2026},
  doi = {10.1140/epjc/s10052-026-15831-9},
  note = {\href{https://doi.org/10.1140/epjc/s10052-026-15831-9}{doi:10.1140/epjc/s10052-026-15831-9}}
}

@article{Nashed2026,
  author = {G. G. L. Nashed},
  title = {{Rotating black holes in perfect fluid dark matter: Horizons, photon orbits, and shadows}},
  journal = {J. High Energy Astrophys.},
  volume = {56},
  pages = {100766},
  year = {2027},
  doi = {10.1016/j.jheap.2026.100766},
  note = {Available online in 2026. \href{https://doi.org/10.1016/j.jheap.2026.100766}{doi:10.1016/j.jheap.2026.100766}}
}

@article{Gralla2019,
  author = {S. E. Gralla and D. E. Holz and R. M. Wald},
  title = {{Black hole shadows, photon rings, and lensing rings}},
  journal = {Phys. Rev. D},
  volume = {100},
  pages = {024018},
  year = {2019},
  doi = {10.1103/PhysRevD.100.024018},
  note = {\href{https://doi.org/10.1103/PhysRevD.100.024018}{doi:10.1103/PhysRevD.100.024018}}
}

@article{Johnson2020,
  author = {M. D. Johnson and others},
  title = {{Universal interferometric signatures of a black hole's photon ring}},
  journal = {Sci. Adv.},
  volume = {6},
  pages = {eaaz1310},
  year = {2020},
  doi = {10.1126/sciadv.aaz1310},
  note = {\href{https://doi.org/10.1126/sciadv.aaz1310}{doi:10.1126/sciadv.aaz1310}}
}

@article{EHTpolar2021,
  author = {{{The Event Horizon Telescope Collaboration}}},
  title = {{First M87 Event Horizon Telescope Results. VII. Polarization of the Ring}},
  journal = {Astrophys. J. Lett.},
  volume = {910},
  pages = {L12},
  year = {2021},
  doi = {10.3847/2041-8213/abe71d},
  note = {\href{https://doi.org/10.3847/2041-8213/abe71d}{doi:10.3847/2041-8213/abe71d}}
}

@article{Perlick2022,
  author = {V. Perlick and O. Y. Tsupko},
  title = {{Calculating black hole shadows: Review of analytical studies}},
  journal = {Phys. Rep.},
  volume = {947},
  pages = {1--39},
  year = {2022},
  doi = {10.1016/j.physrep.2021.10.004},
  note = {\href{https://doi.org/10.1016/j.physrep.2021.10.004}{doi:10.1016/j.physrep.2021.10.004}}
}

@article{Balart2014,
  author = {L. Balart and E. C. Vagenas},
  title = {{Regular black holes with a nonlinear electrodynamics source}},
  journal = {Phys. Rev. D},
  volume = {90},
  pages = {124045},
  year = {2014},
  doi = {10.1103/PhysRevD.90.124045},
  note = {\href{https://doi.org/10.1103/PhysRevD.90.124045}{doi:10.1103/PhysRevD.90.124045}}
}

@article{Bokulic2024,
  author = {A. Bokuli\'c and E. Franzin and T. Juri\'c and I. Smoli\'c},
  title = {{Lagrangian reverse engineering for regular black holes}},
  journal = {Phys. Lett. B},
  volume = {854},
  pages = {138750},
  year = {2024},
  doi = {10.1016/j.physletb.2024.138750},
  note = {\href{https://doi.org/10.1016/j.physletb.2024.138750}{doi:10.1016/j.physletb.2024.138750}}
}

@article{Bokulic2022,
  author = {A. Bokuli\'c and I. Smoli\'c and T. Juri\'c},
  title = {{Constraints on singularity resolution by nonlinear electrodynamics}},
  journal = {Phys. Rev. D},
  volume = {106},
  pages = {064020},
  year = {2022},
  doi = {10.1103/PhysRevD.106.064020},
  note = {\href{https://doi.org/10.1103/PhysRevD.106.064020}{doi:10.1103/PhysRevD.106.064020}}
}

@article{Bokulic2026,
  author = {A. Bokuli\'c and T. Juri\'c and I. Smoli\'c},
  title = {{Conundrum of regular black holes with nonlinear electromagnetic fields}},
  journal = {Phys. Rev. D},
  volume = {113},
  pages = {024044},
  year = {2026},
  doi = {10.1103/z7gd-96ms},
  note = {\href{https://doi.org/10.1103/z7gd-96ms}{doi:10.1103/z7gd-96ms}}
}

@article{Mazharimousavi2026,
  author = {S. H. Mazharimousavi},
  title = {{Physical properties of charged black holes from the nonlinear electrodynamics model based on electric potential regularization}},
  journal = {Eur. Phys. J. C},
  volume = {86},
  pages = {1240},
  year = {2026},
  doi = {10.1140/epjc/s10052-026-16505-2},
  note = {\href{https://doi.org/10.1140/epjc/s10052-026-16505-2}{doi:10.1140/epjc/s10052-026-16505-2}}
}

@article{DeFelice2025,
  author = {A. De Felice and S. Tsujikawa},
  title = {{Instability of nonsingular black holes in nonlinear electrodynamics}},
  journal = {Phys. Rev. Lett.},
  volume = {134},
  pages = {081401},
  year = {2025},
  doi = {10.1103/PhysRevLett.134.081401},
  note = {\href{https://doi.org/10.1103/PhysRevLett.134.081401}{doi:10.1103/PhysRevLett.134.081401}}
}

@article{Tomizawa2023,
  author = {S. Tomizawa and R. Suzuki},
  title = {{Causality of photon propagation under dominant energy condition in nonlinear electrodynamics}},
  journal = {Phys. Rev. D},
  volume = {108},
  pages = {124072},
  year = {2023},
  doi = {10.1103/PhysRevD.108.124072},
  note = {\href{https://doi.org/10.1103/PhysRevD.108.124072}{doi:10.1103/PhysRevD.108.124072}}
}

@article{RussoTownsend2026,
  author = {J. G. Russo and P. K. Townsend},
  title = {{Black holes and causal nonlinear electrodynamics}},
  journal = {J. High Energy Phys.},
  volume = {09},
  pages = {133},
  year = {2026},
  doi = {10.1007/JHEP09(2026)133},
  note = {\href{https://doi.org/10.1007/JHEP09(2026)133}{doi:10.1007/JHEP09(2026)133}}
}

@article{NomuraYoshidaSoda2020,
  author = {K. Nomura and D. Yoshida and J. Soda},
  title = {{Stability of magnetic black holes in general nonlinear electrodynamics}},
  journal = {Phys. Rev. D},
  volume = {101},
  pages = {124026},
  year = {2020},
  doi = {10.1103/PhysRevD.101.124026},
  note = {\href{https://doi.org/10.1103/PhysRevD.101.124026}{doi:10.1103/PhysRevD.101.124026}}
}

@article{Uniyal2024,
  author = {A. Uniyal and S. Chakrabarti and M. Fathi and A. \"Ovg\"un},
  title = {{Observational signatures: Shadow cast by the effective metric of photons for black holes with rational non-linear electrodynamics}},
  journal = {Ann. Phys.},
  volume = {462},
  pages = {169614},
  year = {2024},
  doi = {10.1016/j.aop.2024.169614},
  note = {\href{https://doi.org/10.1016/j.aop.2024.169614}{doi:10.1016/j.aop.2024.169614}}
}

@article{GuzmanMontielBreton2024,
  author = {E. Guzm\'an-Herrera and A. Montiel and N. Bret\'on},
  title = {{Comparative of light propagation in Born-Infeld, Euler-Heisenberg and ModMax nonlinear electrodynamics}},
  journal = {JCAP},
  volume = {11},
  pages = {002},
  year = {2024},
  doi = {10.1088/1475-7516/2024/11/002},
  note = {\href{https://doi.org/10.1088/1475-7516/2024/11/002}{doi:10.1088/1475-7516/2024/11/002}}
}

@article{GrallaLupsascaMarrone2020,
  author = {S. E. Gralla and A. Lupsasca and D. P. Marrone},
  title = {{The shape of the black hole photon ring: A precise test of strong-field general relativity}},
  journal = {Phys. Rev. D},
  volume = {102},
  pages = {124004},
  year = {2020},
  doi = {10.1103/PhysRevD.102.124004},
  note = {\href{https://doi.org/10.1103/PhysRevD.102.124004}{doi:10.1103/PhysRevD.102.124004}}
}

@article{Broderick2022,
  author = {A. E. Broderick and others},
  title = {{How narrow is the M87* ring -- II. A new geometric model}},
  journal = {Mon. Not. R. Astron. Soc.},
  volume = {517},
  pages = {2462},
  year = {2022},
  doi = {10.1093/mnras/stac2743},
  note = {\href{https://doi.org/10.1093/mnras/stac2743}{doi:10.1093/mnras/stac2743}}
}

@article{Bozza2002,
  author = {V. Bozza},
  title = {{Gravitational lensing in the strong field limit}},
  journal = {Phys. Rev. D},
  volume = {66},
  pages = {103001},
  year = {2002},
  doi = {10.1103/PhysRevD.66.103001},
  note = {\href{https://doi.org/10.1103/PhysRevD.66.103001}{doi:10.1103/PhysRevD.66.103001}}
}

@article{BozzaScarpetta2007,
  author = {V. Bozza and G. Scarpetta},
  title = {{Strong deflection limit of black hole gravitational lensing with arbitrary source distances}},
  journal = {Phys. Rev. D},
  volume = {76},
  pages = {083008},
  year = {2007},
  doi = {10.1103/PhysRevD.76.083008},
  note = {\href{https://doi.org/10.1103/PhysRevD.76.083008}{doi:10.1103/PhysRevD.76.083008}}
}

@article{GlampedakisSilva2019,
  author = {K. Glampedakis and H. O. Silva},
  title = {{Eikonal quasinormal modes of black holes beyond general relativity}},
  journal = {Phys. Rev. D},
  volume = {100},
  pages = {044040},
  year = {2019},
  doi = {10.1103/PhysRevD.100.044040},
  note = {\href{https://doi.org/10.1103/PhysRevD.100.044040}{doi:10.1103/PhysRevD.100.044040}}
}

@article{Cardoso2009,
  author = {V. Cardoso and A. S. Miranda and E. Berti and H. Witek and V. T. Zanchin},
  title = {{Geodesic stability, Lyapunov exponents, and quasinormal modes}},
  journal = {Phys. Rev. D},
  volume = {79},
  pages = {064016},
  year = {2009},
  doi = {10.1103/PhysRevD.79.064016},
  note = {\href{https://doi.org/10.1103/PhysRevD.79.064016}{doi:10.1103/PhysRevD.79.064016}}
}

@article{Decanini2011,
  author = {Y. D\'ecanini and G. Esposito-Far\`ese and A. Folacci},
  title = {{Universality of high-energy absorption cross sections for black holes}},
  journal = {Phys. Rev. D},
  volume = {83},
  pages = {044032},
  year = {2011},
  doi = {10.1103/PhysRevD.83.044032},
  note = {\href{https://doi.org/10.1103/PhysRevD.83.044032}{doi:10.1103/PhysRevD.83.044032}}
}

@article{Moncrief1974Even,
  author = {V. Moncrief},
  title = {{Stability of Reissner--Nordstr\"om black holes}},
  journal = {Phys. Rev. D},
  volume = {10},
  pages = {1057},
  year = {1974},
  doi = {10.1103/PhysRevD.10.1057},
  note = {\href{https://doi.org/10.1103/PhysRevD.10.1057}{doi:10.1103/PhysRevD.10.1057}}
}

@article{Moncrief1975,
  author = {V. Moncrief},
  title = {{Gauge-invariant perturbations of Reissner--Nordstr\"om black holes}},
  journal = {Phys. Rev. D},
  volume = {12},
  pages = {1526},
  year = {1975},
  doi = {10.1103/PhysRevD.12.1526},
  note = {\href{https://doi.org/10.1103/PhysRevD.12.1526}{doi:10.1103/PhysRevD.12.1526}}
}

@article{Zerilli1974RN,
  author = {F. J. Zerilli},
  title = {{Perturbation analysis for gravitational and electromagnetic radiation in a Reissner--Nordstr\"om geometry}},
  journal = {Phys. Rev. D},
  volume = {9},
  pages = {860},
  year = {1974},
  doi = {10.1103/PhysRevD.9.860},
  note = {\href{https://doi.org/10.1103/PhysRevD.9.860}{doi:10.1103/PhysRevD.9.860}}
}

@article{Toshmatov2018,
  author = {B. Toshmatov and Z. Stuchl\'ik and J. Schee and B. Ahmedov},
  title = {{Electromagnetic perturbations of black holes in general relativity coupled to nonlinear electrodynamics}},
  journal = {Phys. Rev. D},
  volume = {97},
  pages = {084058},
  year = {2018},
  doi = {10.1103/PhysRevD.97.084058},
  note = {\href{https://doi.org/10.1103/PhysRevD.97.084058}{doi:10.1103/PhysRevD.97.084058}}
}

@article{Okyay2022,
  author = {M. Okyay and A. \"Ovg\"un},
  title = {{Nonlinear electrodynamics effects on the black hole shadow, deflection angle, quasinormal modes and greybody factors}},
  journal = {JCAP},
  volume = {01},
  pages = {009},
  year = {2022},
  doi = {10.1088/1475-7516/2022/01/009},
  note = {\href{https://doi.org/10.1088/1475-7516/2022/01/009}{doi:10.1088/1475-7516/2022/01/009}}
}

@article{Liang2026,
  author = {J. Liang and D. Liu and Z.-W. Long},
  title = {{Quasinormal modes and greybody factors of black holes corrected by nonlinear electrodynamics}},
  journal = {Eur. Phys. J. C},
  volume = {86},
  pages = {17},
  year = {2026},
  doi = {10.1140/epjc/s10052-025-15245-z},
  note = {\href{https://doi.org/10.1140/epjc/s10052-025-15245-z}{doi:10.1140/epjc/s10052-025-15245-z}}
}

@article{Moncrief1974,
  author = {V. Moncrief},
  title = {{Odd-parity stability of a Reissner--Nordstr\"om black hole}},
  journal = {Phys. Rev. D},
  volume = {9},
  pages = {2707},
  year = {1974},
  doi = {10.1103/PhysRevD.9.2707},
  note = {\href{https://doi.org/10.1103/PhysRevD.9.2707}{doi:10.1103/PhysRevD.9.2707}}
}

@article{OuldElHadj2022,
  author = {M. Ould El Hadj and S. R. Dolan},
  title = {{Conversion of electromagnetic and gravitational waves by a charged black hole}},
  journal = {Phys. Rev. D},
  volume = {106},
  pages = {044002},
  year = {2022},
  doi = {10.1103/PhysRevD.106.044002},
  note = {\href{https://doi.org/10.1103/PhysRevD.106.044002}{doi:10.1103/PhysRevD.106.044002}}
}

@article{EHT2019,
  author = {{{The Event Horizon Telescope Collaboration}}},
  title = {{First M87 Event Horizon Telescope Results. VI. The Shadow and Mass of the Central Black Hole}},
  journal = {Astrophys. J. Lett.},
  volume = {875},
  pages = {L6},
  year = {2019},
  doi = {10.3847/2041-8213/ab1141},
  note = {\href{https://doi.org/10.3847/2041-8213/ab1141}{doi:10.3847/2041-8213/ab1141}}
}

@article{EHTSgr2022,
  author = {{{The Event Horizon Telescope Collaboration}}},
  title = {{First Sagittarius A* Event Horizon Telescope Results. VI. Testing the Black Hole Metric}},
  journal = {Astrophys. J. Lett.},
  volume = {930},
  pages = {L17},
  year = {2022},
  doi = {10.3847/2041-8213/ac6756},
  note = {\href{https://doi.org/10.3847/2041-8213/ac6756}{doi:10.3847/2041-8213/ac6756}}
}

@article{EHT2025,
  author = {{{The Event Horizon Telescope Collaboration}}},
  title = {{Horizon-scale variability of M87* from 2017--2021 EHT observations}},
  journal = {Astron. Astrophys.},
  volume = {704},
  pages = {A91},
  year = {2025},
  doi = {10.1051/0004-6361/202555855},
  note = {\href{https://doi.org/10.1051/0004-6361/202555855}{doi:10.1051/0004-6361/202555855}}
}

@article{Tlemissov2025,
  author = {A. Tlemissov and B. Toshmatov and J. Kov\'a\v{r}},
  title = {{Effect of nonlinear electrodynamics on polarization distribution around black holes}},
  journal = {Phys. Rev. D},
  volume = {111},
  pages = {064084},
  year = {2025},
  doi = {10.1103/PhysRevD.111.064084},
  note = {\href{https://doi.org/10.1103/PhysRevD.111.064084}{doi:10.1103/PhysRevD.111.064084}}
}

@misc{EHTdata,
  author = {{Event Horizon Telescope Collaboration}},
  title = {{EHT Data Products}},
  year = {2026},
  note = {release list updated 29 June 2026. \url{https://eventhorizontelescope.org/for-astronomers/data}},
  url = {https://eventhorizontelescope.org/for-astronomers/data}
}

@article{Dahale2025,
  author = {R. Dahale and others},
  title = {{Origin of the ring ellipticity in the black hole images of M87*}},
  journal = {Astron. Astrophys.},
  volume = {699},
  pages = {A279},
  year = {2025},
  doi = {10.1051/0004-6361/202555235},
  note = {\href{https://doi.org/10.1051/0004-6361/202555235}{doi:10.1051/0004-6361/202555235}}
}

@article{BertiCardosoStarinets2009,
  author = {E. Berti and V. Cardoso and A. O. Starinets},
  title = {{Quasinormal modes of black holes and black branes}},
  journal = {Class. Quantum Grav.},
  volume = {26},
  pages = {163001},
  year = {2009},
  doi = {10.1088/0264-9381/26/16/163001},
  note = {\href{https://doi.org/10.1088/0264-9381/26/16/163001}{doi:10.1088/0264-9381/26/16/163001}}
}
\end{document}